\documentclass[]{jfm-stripped}

\graphicspath{{./}}
\usepackage{newtxtext, newtxmath, bm}
\usepackage{pgfplots}

\usepackage{graphicx}
\usepackage{natbib}
\usepackage{hyperref}
\usepackage{ar}
\usepackage[normalem]{ulem}

\title[Automated turbulence modelling]
{A proposal for automated turbulence modelling}

\author[M.~Castelletti, M.~Quadrio]
{Marco Castelletti and Maurizio Quadrio
\corresp{\email{maurizio.quadrio@polimi.it}}
}
\affiliation{
\aff{1} Dipartimento di Scienze e Tecnologie Aerospaziali, Politecnico di Milano,
via La Masa 34, 20156 Milano, Italy
}

\begin{document}

\maketitle

\begin{abstract}
Solving the Reynolds-averaged Navier--Stokes equations (RANS) closed with an eddy viscosity computed through a turbulence model is still the leading approach for Computational Fluid Dynamics simulations. 
Unfortunately, universal models with good predictive capabilities over a wide range of flows are not available. 
In this work, we propose the use of machine learning to improve existing RANS models. The approach is not data-driven, and does not require high-fidelity training data.
A convolutional neural network is used to identify and segment at runtime the flow field into different zones, each resembling one item of a predefined list of elementary flows. 
The turbulence model applied in each zone is taken from an equally predefined set of classic models, each specifically tuned to work best for one elementary flow, free from the requirement of universality.
The idea is first presented in general terms, and then demonstrated via a preliminary implementation, where only three elementary flows are considered, and three turbulence models are used. 
Test cases show that, already in this oversimplified form, automated zonal modelling outperforms the baseline RANS models without computational overhead.
\end{abstract}


\section{Introduction}
\label{sec:introduction}

Artificial intelligence and machine learning (ML) in fluid mechanics have recently seen a blooming interest, and significant progresses are nowadays reported at an impressive pace; even the number of reviews is becoming substantial \citep[see e.g.][for a broad overview]{vinuesa-brunton-2022}. 

In applications involving turbulence, one important contribution from ML is the improvement of closure models for the Reynolds-averaged Navier--Stokes equations (RANS). 
In RANS, turbulence models are required to close the mean-flow equations. The Reynolds stresses tensor, which ensues after the flow variables undergo the Reynolds decomposition and the Navier--Stokes equations are averaged in time, must be replaced by a closed expression involving mean quantities only. A similar requirement holds for the scalar fluxes.
Most existing turbulence models, widely used in the so-called Computational Fluid Dynamics (CFD) realm, resort to the concept of eddy or turbulent viscosity, attributed to Bousinnesq, where a constitutive relation is assumed to exist for the turbulent flow; this relation is written to mimic that of a viscous fluid, and builds upon a scalar quantity, the eddy viscosity.
Other models, known as Reynolds stress transport models (RSTM), compute each component of the Reynolds stress tensor independently, via a dedicated partial differential transport equation where several terms need to be modelled.
While RSTM models, not bound by the Bousinnesq hypothesis, might in principle outperform eddy-viscosity models, the latter are often preferred in applications, owing to their superior simplicity, robustness and appeal in terms of computational cost. They represent the current {\em de facto} standard for industrial CFD simulations, and are becoming the limiting factor in CFD: continued progresses in numerical analysis and computer hardware are such that numerical errors for steady-state flow solutions, even for complex geometries, are almost negligible. Errors associated with turbulence modelling, though, remain finite, and often too large.

Turbulence models have received extensions, improvements and patches over the years, and nowadays they work acceptably well out of the box in most applied flow problems. However, ample room for improvement still exists in terms of accuracy and robustness of the results. 
For example, RANS models contain empirical coefficients, whose values are determined such that the model yields reasonable predictions across a wide range of flows. The specific way these values are calibrated varies; \cite{pope-2000} in his book explains how they are set  such that the model is correct or nearly correct when applied to certain representative flows, where only one or few coefficients matter, and either an exact solution is available, or the correct behavior is empirically known. 
By design, then, any set of values for the coefficients of the typical model represents a compromise; the expert CFD practitioner is aware that tweaking the default values (either globally or locally) might improve the overall performance of the model for a specific flow.
It is also known that each model carries its own strengths and weaknesses, a concept once again well illustrated by \cite{pope-2000}: for example, the popular $k-\epsilon$ model \citep{launder-spalding-1974} has troubles near solid walls, whereas the $k-\omega$ model \citep{wilcox-1988} suffers from the large sensitivity to the uncertain boundary conditions in the free-stream. 
Choosing the turbulence model that best suits the flow of interest is a difficult step in the workflow of a CFD simulation, and a skill not possessed by everybody.

ML-based approaches are quickly changing this picture. 
An excellent review specifically devoted to the ML contribution to turbulence modelling is provided by \cite{duraisamy-iaccarino-xiao-2019}. They describe how modelling could rely upon and be driven by data. Techniques are available to perform statistical inference for assimilation of large quantities of reliable data, and ML can be used to produce a suitable mapping between such large data sets and the quantities of interest. 
Notable among the several examples is the work by \cite{ling-kurzawski-templeton-2016} and \cite{ling-etal-2016}, where a random forest, later upgraded to become a tensor basis neural network that guarantees Galilean invariance, is used to predict the anisotropy of the normalized Reynolds stress tensor. The model was tested in a duct and for the flow over a wavy wall, with encouraging results towards a data-driven modelling of turbulence. A similar approach was adopted by \cite{kaandorp-dwight-2020}. 
\cite{wang-wu-xiao-2017} took a different path, and used available high-accuracy training data to build functions of the discrepancies between the RANS-predicted and the true Reynolds stresses. These functions would then be used to correct the baseline RANS predictions of the Reynolds stresses in flows for which training data are not available.
\cite{sun-etal-2022} proposed the use of a neural network to replace the Spalart--Allmaras (SA) turbulence model \citep{spalart-allmaras-1992} by directly mapping flow features into an eddy viscosity, and demonstrated the coupling of a neural network with a CFD solver. 

The future of ML-based turbulent models, though, is not unanimously seen as bright.
Review papers exist, as the ones by \cite{rumsey-coleman-2022}, \cite{spalart-2023} and \cite{girimaji-2024}, whose outlook is only partially optimistic. They observe that, to date, no general-purpose ML-based model has been produced, much less found to be successful. 
The data-driven approach is indeed plagued by the fundamental issue that a large amount of reliable, high-fidelity data must be available to train the model. The minimum size of the dataset required for a good training of a universal turbulence model is difficult to estimate. Moreover, such data are expensive to produce, either numerically or experimentally, in particular for complex geometries and high values of the Reynolds number. 
Relying on data alone to substitute the turbulence model might also lead to a lack of explainability, especially when deep learning techniques are used. Additional issues concern practical yet fundamental aspects, like e.g. how to make a ML model usable by the CFD community.

Instead of using ML to create new RANS models, in this paper we describe an alternative route where ML is used to make a better use of existing RANS models. 
The importance of developing new models and improving the existing ones is fully acknowledged. However, a new emphasis is given here to the untapped potential of existing eddy-viscosity models, by specifically addressing their lack of universality.
The basic idea behind our approach stems from the observation that existing models can already be used to successfully predict complex flows, provided the flow field can be separated into zones with distinct physical structures, and models are properly tuned to work well for each flow zone.
This immediately reminds of the concept of the so-called zonal modelling \citep{kline-1980, avva-kline-ferziger-1988}, where distinct turbulence models are applied to those portions of the flow where each is supposed to work at its best. By relaxing the constraint that the same model (or, in alternative, the same values for the coefficients of one model) should be used in one simulation, zonal modelling was shown to perform interestingly well. Its obvious drawback is the required {\em a priori} knowledge of the solution, necessary to assign different models to different zones. To quote \cite{spalart-2023}, "the idea of a zonal model explicitly controlled by the user to optimize it in each area is very unattractive in industry". 
The present approach, dubbed automated turbulence modelling or ATM, precisely addresses this drawback.

This is not the first time that zonal modelling is reformulated to take advantage of ML.
For example \cite{matai-durbin-2019-2} successfully introduced a scalar corrective function $\beta$ in front of the production term in the $\omega$ equation of the $k-\omega$ model, and used a neural network to predict $\beta$ starting from the flow variables, thus in fact arriving at a zone-dependent modelling of turbulence.
\cite{wu-zhang-zhang-2024}, using a conditioned field inversion technique, and \cite{aulakh-yang-maulik-2024}, relying on ensemble Kalman filtering, attempted a calibration of the default values of the model coefficients, and obtained improved predictions of separated flows without detrimental effects for attached flows.
\cite{dezordo-etal-2024} developed a space-aggregated turbulence model to predict the flow through a compressor cascade. Predictions from multiple RANS models were linearly combined using a random-forest algorithm to produce an aggregate solution that can in principle outperform those of the individual models. 
In general, however, high-fidelity data are always necessary to train ML-based zonal turbulence models. They were produced via large-eddy simulations (LES) by \cite{matai-durbin-2019-2} and \cite{wu-zhang-zhang-2024}, and via experiments by \cite{aulakh-yang-maulik-2024}. The random forest used by \cite{dezordo-etal-2024} requires high-fidelity data too; however, lacking such data, in that proof-of-concept study reference data generated with an RSTM model were used instead.
The works by \cite{lozano-bae-2023} and \cite{arranz-etal-2024} concern LES models instead of RANS models. However, they are worth mentioning here, since they share with the present approach the basic philosophy of identifying elementary flows. An LES wall model is built as a combination of building blocks, under the assumption that a small number of simple flows, such as wall-bounded laminar and turbulent flows or statistically unsteady wall turbulence, contain the essential physics required to formulate generalizable LES wall models. Their approach is implemented using two neural networks: a classifier and a predictor. The classifier is fed data from the LES solver and quantifies similarities between the input and the collection of building-block flows; the predictor generates the wall-shear stress prediction via a combination of the building-block flows from the database.

The zonal automated turbulence model (ATM) introduced in this paper targets RANS simulations, does not require high-fidelity training data, and uses ML to improve the traditional zonal modelling in the crucial selection of flow zones, for which no user input is required.
To set up ATM, a (short) dictionary of structural or elementary flows is preliminarily created, where each flow is simple enough to allow one model to work well. Different elementary flows can be dealt with by different models, or by the same model using different values of its empirical constants. 
Once ATM is set up, during a generic CFD simulation, a neural network designed for the semantic segmentation of the flow field into the various elementary flows, i.e. the "words" of the dictionary, examines the solution at runtime, and identifies the flow zones; each zone is then assigned to its optimal RANS model.
This approach works around the need for large amounts of high-fidelity data, and exploits flow physics to improve the model performance; the traditional eddy-viscosity models, with their accompanying wealth of information accumulated by the community during half a century of intense CFD practice, are not trashed but used at the best of their predictive power.

This paper provides an overview of the ATM procedure, and describes a preliminary, proof-of-concept implementation, in which the complexity is kept to a minimum. Only flows that are two-dimensional in the mean are considered, only three elementary flows are defined, and correspondingly only three RANS models are employed. 

The structure of this paper is as follows. After this introduction, \S\ref{sec:overview} provides an overview of the approach, and \S\ref{sec:segmentation} describes with some detail the segmentation network. The implementation of the ATM procedure into an existing CFD solver is presented in \S\ref{sec:models}. The performance of the segmentation network is described in \S\ref{sec:results-segmentation}, whereas \S\ref{sec:results-model} addresses the predictive capabilities of ATM applied to two simple test cases (the flow over a backward-facing step and the flow over a bump), one of them entirely unseen by the segmentation network during training. Finally \S\ref{sec:conclusions} presents a concluding discussion.

\section{ATM: an overview}
\label{sec:overview}

With ATM, the flow field computed during a RANS simulation is segmented at runtime to automatically identify structural flow zones. Each zone is put in correspondence to a turbulence model, previously optimized to deal with the physics of the flow in that particular zone.
This section describes the main logical steps of the ATM procedure.

\subsection{A dictionary for semantic segmentation}
A structural flow zone is a portion of the flow field that contains a certain flow feature. 
Attempts at the compilation of comprehensive lists of flow features/zones exist in the literature.
A relatively small set of 20 canonical flows was identified by \cite{kline-1980}, by including two- and three-dimensional attached boundary layers, wakes and jets, reattachment and detachment regions, recirculations, secondary flows, transitional flows.
To make the list suitable for zonal modelling, \cite{avva-kline-ferziger-1988} reduced its cardinality to three, by retaining curved streamlines, free-shear layers and attached boundary layers. 

Alternative choices for creating the zone list are possible. For example \cite{ling-kurzawski-templeton-2016} trained the neural network employed for enforcing Galiean invariance to their model by using nine canonical flows, including flows over planar and non-planar walls and jets in cross-flow.
Choosing the structural flow zones to be included in the dictionary is a crucial first step that would, in principle, deserve a study on its own. Each zone should be meaningful and general, while remaining elementary enough that a single turbulence model could be properly tuned to work very well on it. An interesting point that is worth stressing in the present context is that the dictionary of flow zones can be easily extended once the importance of a missing elementary flow is recognized.

\subsection{Semantic segmentation}
\label{sec:semantic-segmentation}

Once a dictionary with the list of canonical zones is created, a ML algorithm in the form of  a neural network is trained for the semantic segmentation of a generic flow field into zones. 
A neural network (hereinafter abbreviated with NN) is a model, designed to mimic the human brain, that uses artificial neurons to process data and perform tasks such as image and video recognition, natural language processing, and time series analysis.
Each neuron receives input from other neurons or external sources, passes this information through an activation function, and sends its output to other neurons. 
Each neuron is characterized by a set of weights and bias, optimized in the so called training process. 
The precise network structure is defined by the choice of its hyperparameters. 

In our case, the NN is given a segmentation task to be carried out at runtime. Semantic segmentation is a computer vision task that associates a class label to each pixel of an image; it is used to recognize collections of pixels belonging to distinct categories, without separating distinct instances of the same class.
The mean flow field computed by CFD is interpreted as a two- or three-dimensional pseudo-image (i.e. a data structure with a two- or three-dimensional array), which constitutes the input to the NN. The algorithm associates each pixel or voxel of the pseudo-image to one of the labels contained in the dictionary.
The values of velocity components and pressure (an, possibly, of other mean flow quantities) are organized into a corresponding number of channels of the pseudo-image, by sampling the flow variables uniformly in space. The spatial information is conveyed by the position in the input matrix. The output of the NN is, for each pixel, the probability to belong to each label; the pixel label is then selected as the label with the highest probability. 

The segmentation network must be preliminarily trained in a supervised way, i.e. by solving an optimization problem that computes values of weights and biases by minimizing the difference, quantified via a loss function, between the predicted output and a known ground truth.
A training dataset is therefore required. Such a dataset must of course be sufficiently general, i.e. it must contemplate all the flow zones present in the dictionary. 
A key feature of the ATM training procedure is that high-fidelity data are not required: the network only needs to learn the qualitative features of the flow zones. In practice, standard RANS simulations or even laminar flow fields could suffice for training; in the present work, as described in \S\ref{sec:dataset}, only laminar flow solutions are used for training.

It should be noted that a supervised NN is not the only option to solve the flow field segmentation task using ML. Other approaches are available, such as unsupervised learning methods, that can accurately identify flow regions without requiring ground truth. A notable example is the recent work by \cite{saetta-tognaccini-2023}, where an unsupervised Gaussian mixture algorithm is used to detect grid cells belonging to three different classes (boundary layer, shock wave, and inviscid flow) in RANS simulations of the flow around a wing.

\begin{figure}
\centering
\includegraphics[width=\textwidth]{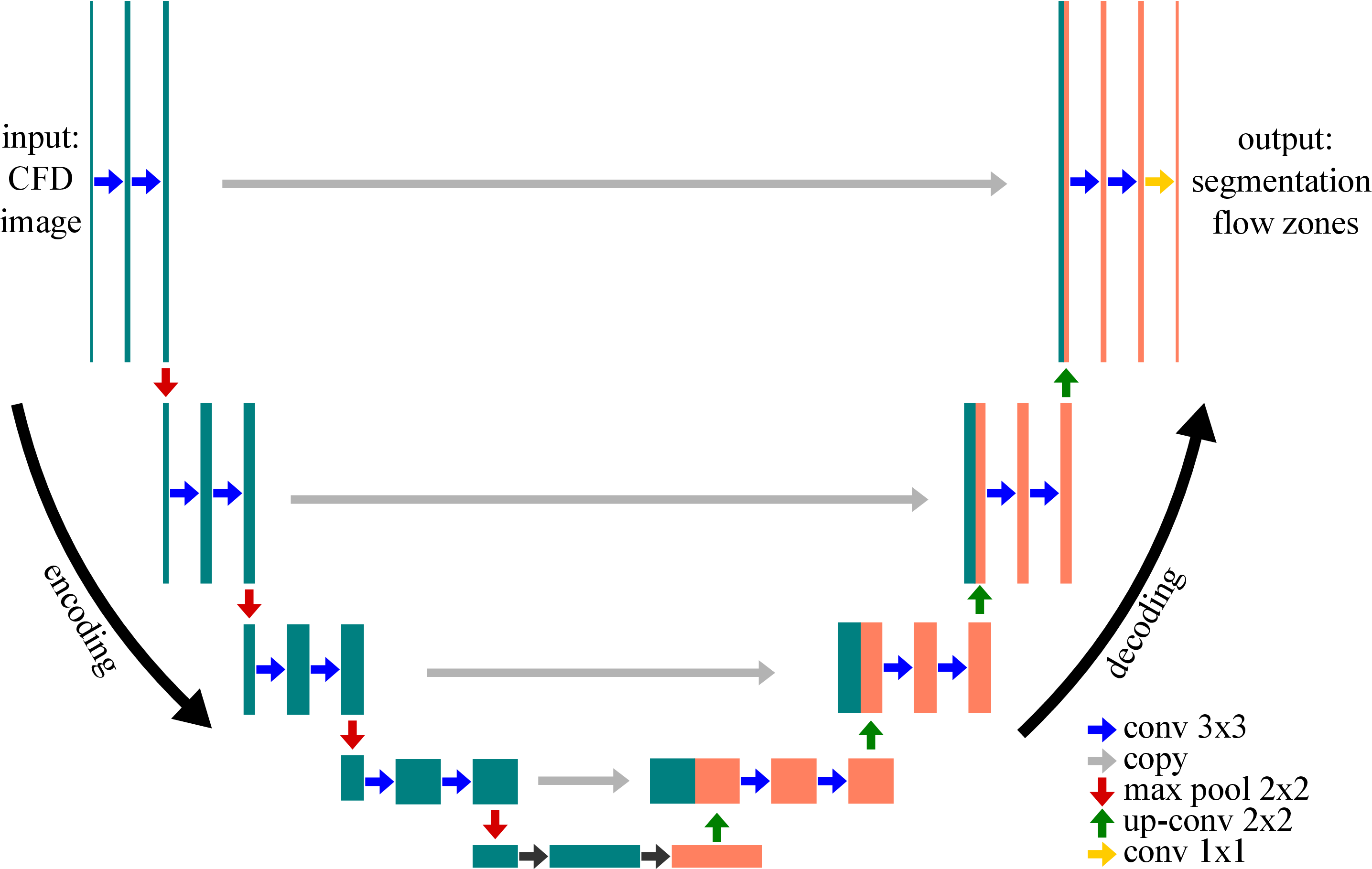}
\caption{The conceptual architecture of the U-net, adapted from \cite{ronneberger-fischer-brox-2015}. For simplicity, the CFD image in input is assumed one-dimensional.}
\label{fig:unet}
\end{figure}

The semantic segmentation paradigm is well established within the computer vision community, where a modern convolutional segmentation network is the U-net, introduced by \cite{ronneberger-fischer-brox-2015}, originally intended for biomedical applications. 
A convolutional network suits well the present application, because of its ability to easily process input of every size, an important feature if one considers the variety of domain sizes encountered in CFD. 
The U-net architecture is schematically illustrated in figure \ref{fig:unet}. 
For graphical convenience, the input image in the top left is assumed to be one-dimensional, so that the vertical size of the green bars corresponds to the number of pixels; their horizontal size reflects the number of channels (or feature maps) of the input image. 
By first applying a predefined number of filters with a spatial size of $3 \times 3$ pixels through convolutions (blue arrows), the U-net increases the number of channels of the image, which is then downsampled through a cascade of max-pooling layers (red arrows), where the spatial resolution is reduced by retaining maximum values over $2 \times 2$ pools of pixels, extracting key features. Downsampling implies that the height of the green bars is reduced. 
Through the encoding process, which takes place on the left side of the figure, semantic content is extracted from the image by condensing spatial information while increasing the number of channels. 
After the encoding, i.e. in the right portion of the U, decoding takes place (orange rectangles); up-convolutional layers (green arrows), with filters of spatial size of $2 \times 2$, progressively restore the spatial resolution, while the number of channels decreases.
Via a last convolution layer, with filters of $1 \times 1$ size, the network eventually produces a segmentation mask for the original image, where the number of channels corresponds to the number of output labels.
Since the deepest layers of the encoder contain limited spatial information, the U-net architecture also incorporates skip connections (grey arrows), that copy high-spatial, low-semantic information from the encoder and concatenate it with the low-spatial, high-semantic data of the decoder.

\subsection{Turbulence model(s)}
The third logical step consists in associating each of the structural flow zones to a turbulence model, previously calibrated to work at its best for that particular elementary flow.
In practice, more than one approach can be envisaged. Each flow zone can be assigned a specific turbulence model, or a combination of a certain model and a set of coefficients. Of course, one could decide to work with the same model through all the zones, and just vary the coefficients zone-wise.
To minimize the difficulty of blending the formulation across regions with different labels, it is advisable that the models for the various zones use a coherent mathematical formulation. Choosing models of the eddy-viscosity type, for example, naturally tends to yield a seamless transition from zone to zone, as seen in \cite{avva-kline-ferziger-1988} and confirmed here.

The ATM formulation (which includes the dictionary, the set of model/coefficients and the trained NN that provides segmentation at run time) must be then integrated with a CFD solver. This can be cumbersome to a variable degree, depending on the architectural choices taken above. Integration is immediate whenever the same model with spatially variable coefficients is used, or ATM is based on a model where one or more tunable parameters/functions exist, as in the $k-\omega$ SST model \citep{menter-1993} where one free function is available, or in the generalized $k-\omega$ model GEKO \citep{menter-lechner-matyushenko-2022}, where the free functions are three. 
Access to the solver source code is essential when fine-tuning or customization of the ATM model is required. 
It will be shown later that the details of the integration between ATM and the RANS solver are not critical, and that the iterative solution process of the CFD solver consistently leads to the same solution.

\subsection{Our implementation}
In this work, ATM is presented via an extremely simplified implementation. 
First of all, our dictionary of elementary flows only contains two-dimensional mean flows.
The list of 20 canonical flows proposed by \cite{kline-1980} is shortened, and only three of them are retained in the segmentation dictionary, thus reducing the cost of creating a preliminary training dataset.
The selected zones, according to the classification proposed by \cite{kline-1980}, are: zone n.2 (two-dimensional, attached boundary layer), zone n.5 (mixing layer) and zone n.10 (recirculation). 
As done later by \cite{avva-kline-ferziger-1988} in their zonal-model experiments, zone n.5 has been further restricted to only include free-shear layers above separated regions. 
Details on how  the segmentation is carried out are provided in \S\ref{sec:segmentation}.

As for the turbulence model, in this preliminary work we opt for putting together three turbulence models, namely the well known $k-\omega$ and $k-\epsilon$ models and a modified version of $k-\omega$ which features a rotation/curvature correction. They are considered in their pristine form, with default values for their coefficients, which are therefore constant across the computational domain. In other words, just a a solution-based blending of the three models is considered, akin to $k-\omega$ SST. Details on the turbulence models are provided in \S\ref{sec:models}. 

\section{The segmentation network}
\label{sec:segmentation}

In the present implementation, the list of canonical flow zones introduced by \cite{kline-1980} only includes "two-dimensional attached boundary layers" (label {\em BL}), "recirculation" (label {\em R}), and "free-shear layers above recirculation" (label {\em SL}). 
Two additional words are contained in our dictionary, namely "solid body" (label {\em X}) and "featureless flow" (label $\circ$). In the end, the dictionary contains 5 words and a corresponding number of labels. 

\subsection{The neural network}

The U-net architecture described above in \S\ref{sec:semantic-segmentation} in general terms is now simplified and specialized to the case of interest.
Only two-dimensional flows are considered. The input image possesses three channels (two components of the mean velocity and pressure); however, we have empirically determined that, for the extremely small word set contained in the present dictionary, pressure can be safely removed from the input, without detrimental effects on the classification accuracy, thus leading to two-channels images. 
This is certainly expected to change as soon as the complexity of the ATM framework is increased. For example, besides introducing the pressure channel, it might be useful to add turbulence quantities to aid the classification process. It is important to mention that the network architecture is readily adjustable for upscaling. Although only five labels are considered here, their number can be increased by simply increasing the number of filters used in the convolutional layers. Also, extending the procedure to three-dimensional flows is straightforward from the NN perspective.

The U-net algorithm used in this work is written in Python, using the TensorFlow library \citep{abadi-etal-2016}.
Our U-net has the same number and type of layers as the original U-net by \cite{ronneberger-fischer-brox-2015}, and therefore possesses more than $5 \cdot 10^6$
trainable parameters, optimized during the training process. 
All the internal layers of the NN use the ELU activation function \citep{clevert-unterthiner-hochreiter-2015}; the last output layer employs the sigmoid activation function, which outputs a probability for each of the output labels.
Segmentation networks have seen limited use in fluid mechanics to date, although some recent studies have employed segmentation to isolate vortical regions in the flow. 
\cite{kashir-etal-2021} used an U-net to detect vortex structures in a lid-driven cavity flow. A similar network was employed by \cite{deng-etal-2019} and \cite{deng-etal-2022} to detect vortices around bodies of different shapes. 
\cite{strofer-etal-2019} employed three distinct networks designed to identify vortical structures, boundary layers and regions of flow recirculation.

\subsection{The training dataset}
\label{sec:dataset}

\begin{figure}
\centering
\includegraphics[width=0.8\textwidth]{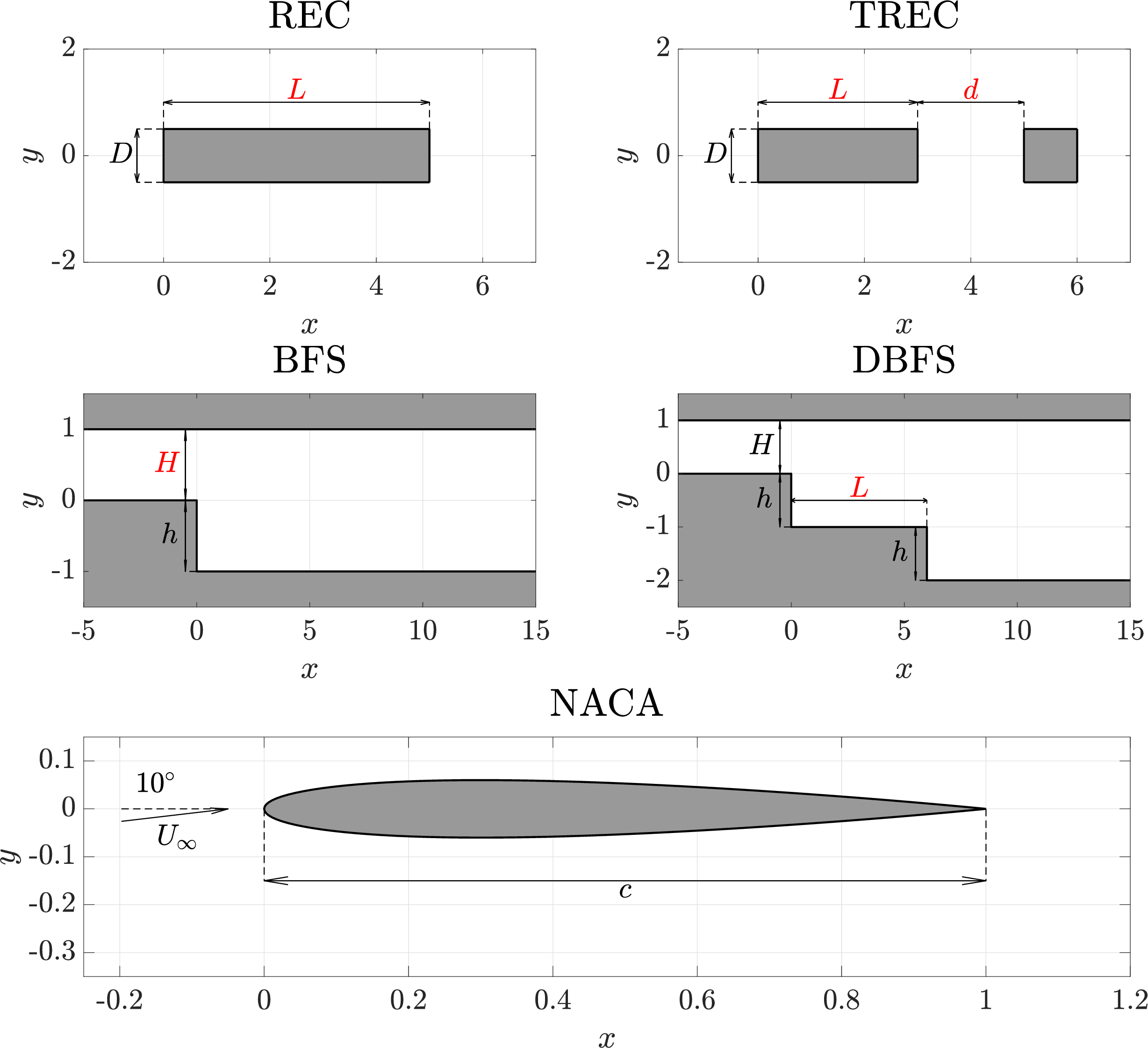}
\caption{Sketch of the five flow cases used for in the training stage: rectangle (REC), two rectangles in tandem (TREC), a backward-facing step (BFS), a double backward-facing step (DBFS), and a NACA airfoil at incidence (NACA). The main geometrical parameters are indicated; quantities in red are parametrically varied to augment the dataset.}
\label{fig:geometries}
\end{figure}

A dataset of CFD-computed flow fields is created on purpose for training the U-net. 
Simple numerical simulations are carried out for several two-dimensional flows encompassing the entire word set in the dictionary. The considered flows include single or multiple bluff bodies immersed in an uniform flow, a single and double backward-facing step, and a NACA 0012 airfoil at incidence.
This set of flows is merely a starting point, yet both internal and external flows are included, and bluff and aerodynamic bodies as well; all cases present separations and recirculations, a free-shear layer above them, and an attached boundary layer. 

Solutions are computed numerically by solving the steady, laminar, two-dimensional incompressible Navier--Stokes equations with the finite-elements code FreeFEM++ \citep{hecht-2012}, after discretization on an unstructured grid. 
The equations are solved in their non-dimensional form, so that results can be fed directly to the NN without normalization. 
The iterative solution, obtained with the Newton method, is computationally cheap: each case requires a computing time of the order of seconds (on a single core of a personal computer).
Using a laminar dataset in a work on turbulence modeling may seem contradictory at first. However, the qualitative characteristics of the mean turbulent flow are similar to those of their laminar counterparts (a feature that is not unrelated to the success of the eddy viscosity concept); using laminar flows for training is just more efficient, and allows a faster generation of the dataset. 
At any rate, in \S\ref{sec:results-segmentation} the ability of the NN to segment turbulent flow fields will be demonstrated. Nothing precludes to add turbulent cases in future to improve the training of the NN, should additional structural flow zones require a true turbulent field at the training stage.

The flows used for training are sketched in figure \ref{fig:geometries}. The first is a rectangular body (REC), of cross-stream size $D$ (the reference length), immersed in a uniform flow $U_\infty$; the aspect ratio $\AR$ varies in the range $1 \le \AR \le 10$. For the two rectangles in tandem (TREC), the front one has $1 \le \AR \le 8$ but the rear one is always a square, with the distance $d$ separating the two bodies varying from $d/D=1$ to $d/D=6$. Here the Reynolds number is defined as $Re = U_\infty D / \nu$, where $\nu$ is the kinematic viscosity of the fluid. 
The computational domain for REC extends for $-25D < x < 50D$ in the streamwise direction and for $-40D < y < 40D$ in the vertical direction; for TREC, the streamwise length is increased to $-25D < x < 70D$, and the cross-stream length is $-25D < y < 25D$. 
The mesh consists of 40,000 elements for REC, and of 60,000 elements for TREC. $Re$ is varied between 10 and 400.

Another family of flows involve a plane channel flow with one or two steps on one wall, causing a sudden expansion. The height $h$ of the step is the reference length, kept fixed in all cases. For the single backward-facing step (BFS), two values of the height $H$ of the upstream channel are used, namely $H=h$ and $H=0.5h$. 
For the double step (DBFS), only the distance $L$ between the steps is varied, in the range $4h \le L \le 20h$, with $H = h$. The Reynolds number is defined here with the length scale $h$ and the velocity scale made by the bulk velocity at the inlet. The streamwise length of the computational domain is $72h$ for BFS, and varies from $64h$ to $80h$ for DBFS. The number of mesh elements varies between 35,000 (BFS) and an average value of 50,000 for DBFS. $Re$ is varied between 50 and 1300.

Lastly, for the NACA 0012 profile (NACA), only the Reynolds number (defined with the profile chord $c$ and the incoming velocity $U_\infty$) is varied between the values 100 and 4,000. The computational domain is a rectangle, extending for $-7c < x < 13c$ in the streamwise direction and $-8c < y < 8c$ in the vertical direction; approximately 60,000 elements are used. 

Figure \ref{fig:geometries}, where the various cases are sketched, draws in red the geometrical parameters that are varied across the dataset. Additionally, the Reynolds number too is varied (while remaining below the critical value of convergence for the Newton method). In total, 40 simulations have been run for REC, 40 for TREC, 26 for BFS, 36 for  DBFS, and 24 for NACA. 
The main flow features vary with $Re$ and with the geometrical parameters. For example in BFS and DBFS a recirculation bubble develops after the step, and grows as $Re$ increases. Also, a secondary recirculation appears on the upper wall, but only at large $Re$. Similarly, REC always shows a main recirculation after the trailing edge, but the boundary layer may detach also along the horizontal sides, especially at high $Re$ and $\AR$. 


\subsection{Training the network}
\label{sec:training}

\begin{figure}
\centering
\includegraphics[width=0.8\textwidth]{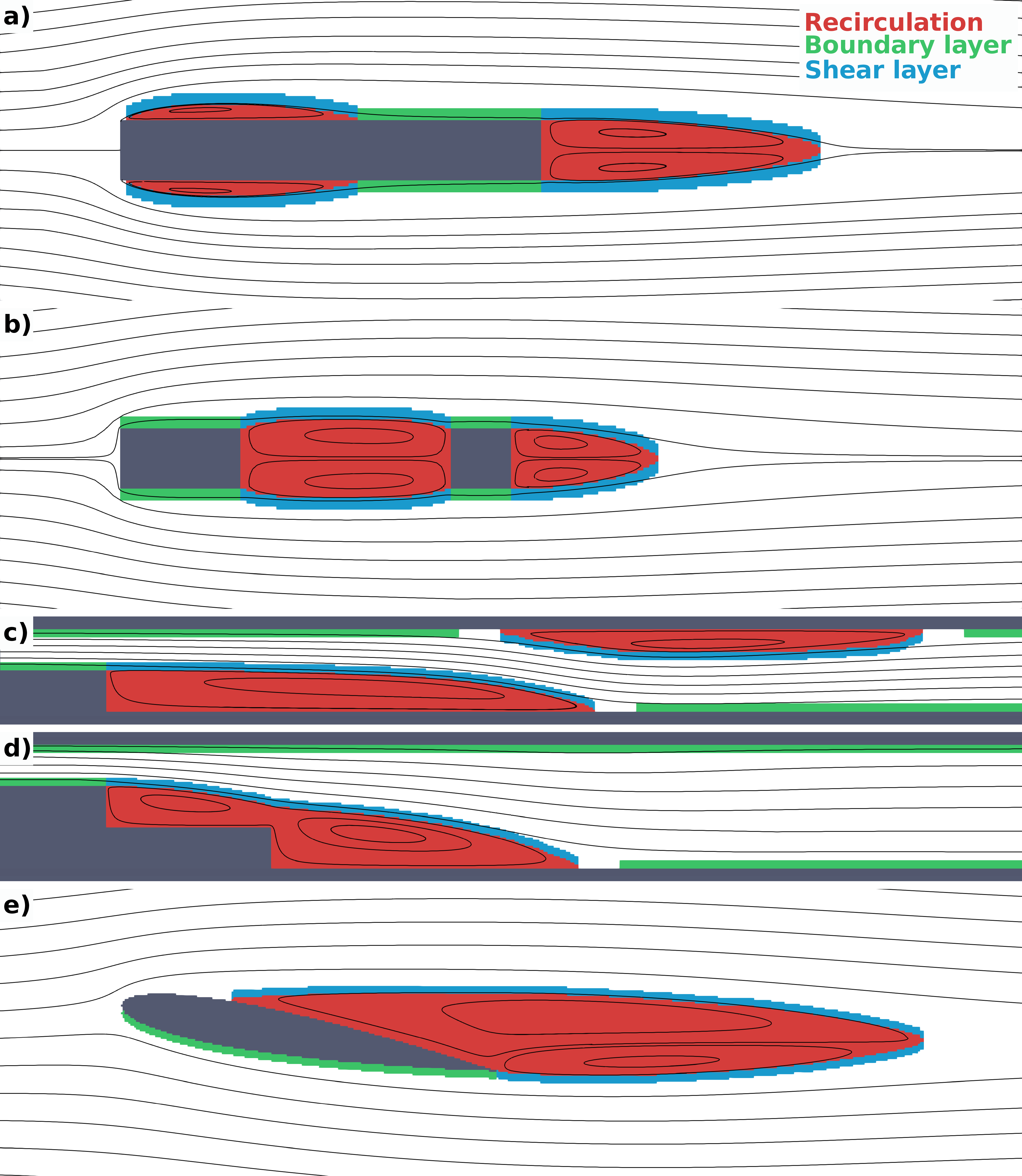}
\caption{Example of manual ground truth annotations of laminar solutions. In each panel, the solid body (label {\em X}) is drawn in gray, and white background indicates featureless flow (label $\circ$). Red indicates a recirculation (label {\em R}), green indicates the attached boundary layer (label {\em BL}) and light blue indicates a free-shear layer above recirculation (label {\em SL}). Panel (a): REC, for $\AR = 7$ and $Re = 240$; panel (b): TREC, front rectangle with $\AR = 2$, $d/D = 5$ and $Re = 100$; panel (c): BFS, for $H/h = 1$ and $Re = 600$; panel (d): DBFS for $L/h = 2$ and $Re = 150$; panel (e): NACA, for $Re = 1600$.}
\label{fig:dataset}
\end{figure}

After the training dataset set is created, the corresponding ground truth must be defined for every image of the dataset. 
In line with the preliminary nature of the procedure, the ground truth is obtained manually, by associating each pixel of the input images to the corresponding labels. 
Since the geometry is known, the label {\em X} of the solid is easily determined. 
The three main labels, i.e. {\em BL, R, SL}, referring to boundary layer, recirculation, and shear layer, are determined by visual inspection and manual marking of every image in the dataset. In particular, label {\em R} is assigned based on the observation of the streamlines; labels {\em BL} and {\em SL}, instead, are annotated by looking at the velocity and its gradient, respectively near the wall and at the edge of recirculating regions.
The fifth label, i.e. the featureless background, is assigned by exclusion to those pixels that eventually do not receive any other label. Examples of manual annotation are reported in figure \ref{fig:dataset}.

The manual annotation is the most time-consuming step of the procedure. While some of its parts are trivial, the time needed to manually segment an image is still of the order of minutes. No particular attention has been paid to streamline the process, and strategies for better automation definitely exist. 
However, ground-truth annotation is carried out only once, and the total number of images required for proper training, which amounts to 166 in our preliminary implementation, would remain limited even in a full-scale model.

After annotation, the dataset is divided into two subsets: a training set and a test set. The training set is used to optimize the parameters of the model; the test set is used to assess the performance of the NN by testing it on never-seen-before data. 
Since five flow configurations are available, five distinct NN are trained. For each, four configurations are used for training, and the remaining one is left out for testing. 
Such training allows to better understand the results of the training process, particularly when the dataset is small. 
During inference, the average of the results from all the networks is used, to minimize noise and obtain more robust predictions. 
The five networks have been trained on an NVidia Tesla GPU P100 (16 GB), with a process that has required a computing time of about three hours. Once again, it is worth noting that training is a one-time occurrence.

Among the various hyperparameters of the NN, the one that is worth discussing here is the loss function $L$, i.e. the function that is minimized during training. $L$ is a metric that quantifies the distance between the predicted output and the correct output. 
A commonly used loss function for multiclass classification problems is the categorical cross entropy $L_{CCE}$ \citep{deng-etal-2022}. It is defined as:
\begin{equation*}
  L_{CCE} = -\sum_{j=1}^{N_p} \sum_{i=1}^{N} t_{ij} \log s_{ij} ,
\end{equation*}
where the index $j$ spans the total number of pixels $N_p$ in the image. The index $i$, instead, spans the total number of labels $N$ to be predicted by the NN (in our case $N=5$). $t_{ij}$ is the ground-truth value for class $i$ at pixel $j$, i.e. $t_{ij}=1$ if the pixel $j$ belongs to class $i$, and $t_{ij}=0$ otherwise. The quantity $s_{ij}$, instead, is the probability (hence, a quantity that varies between 0 and 1) predicted by the NN for the  pixel $j$ to belong to class $i$. The minus sign in front of the formula makes $L_{CCE}$ a definite positive quantity, which is a required property for a function to be minimized during training.

In the present case, simply using $L_{CCE}$ as a loss function does not provide satisfactory results: since a significant portion of image pixels carry the label $\circ$ (featureless flow), the NN ends up being biased towards this label. 
Therefore, the loss function must be improved via an additional term, namely the intersection over union, or $IoU$ \citep{garcia-etal-2017}. The $IoU$ metrics is often used in segmentation tasks; it is a real number, between 0 and 1, defined for each label $i$ as:
\begin{equation}
  IoU_i = \frac{T_i\cap S_i}{T_i\cup S_i}
\label{eq:iou}
\end{equation} 
where $T_i$ stands for the ground-truth label mask, and $S_i$ for the predicted mask for label $i$. 
In words, for each label $IoU$ is computed as the ratio between the number of pixels where prediction and ground truth agree, and the total number of pixels.
In principle, $IoU$ as expressed by Eq.\eqref{eq:iou} is not differentiable and cannot be therefore used as a loss function; however, an equivalent but differentiable metrics can be built with minor tweaks \citep{rahman-wang-2016}:
\begin{equation}
    IoU_i = \frac{\sum_{j=1}^{N_p} T_{ij} S_{ij}}
                 {\sum_{j=1}^{N_p} (T_{ij} + S_{ij} - T_{ij} S_{ij})}
\end{equation}
where all operations are performed pixel-wise.
Since the optimal training must correspond to a minimum of the loss function, the final form for the $IoU$ loss function is $L_{IoU} = \sum_{i=1}^{N} 1 - IoU_i$.
The overall loss function to be minimized during training is the sum of $L_{CCE}$ and $L_{IoU}$:
\begin{equation}
    L = L_{CCE} + L_{IoU} .
\end{equation}

We have found that the $L_{CCE}$ is quite effective at the beginning of the training process, when the predictions are still inaccurate. At later stages, however, $L_{CCE}$ becomes orders of magnitude smaller than $L_{IoU}$. Therefore, only the combination of the two loss functions works acceptably well, while neither alone leads to an effective training.

During training, it is important to avoid the problem of overfitting, i.e., learning too much from the training data and losing the ability to generalize to new data. 
To tackle overfitting, we rely on a technique known as spatial dropout \citep{tompson-etal-2015}: after the convolutional layers of the encoder side of our U-net, during training entire feature maps are randomly dropped, instead of individual neurons, thus improving robustness and generalizability of the model.

Further, a simple data augmentation strategy is put in place. It consists in adding extra pseudo-images, selected randomly during each epoch of the training phase, where the sign of the $x$ velocity component is flipped. 
In this way, the NN learns to avoid predicting labels solely based on the sign of the velocity components, and to discriminate between a recirculation region and a region with negative $x$ velocity component, therefore improving the ability of the network to generalize.

 
A last problem worth mentioning here is the large variability in the size and shape of the input pseudo-images resulting from the numerical simulations. For example, for the REC cases the computational domain is almost a square, while for BFS and DBFS it is a rectangle with a large aspect ratio.
Variability of input size is a common problem in computer vision, where images need to be of the same size, at least within the same batch, to facilitate image loading on the GPU. In fact, conventional data generators are designed to process only images of the same size. Usually the problem is dealt with by stretching the image along one axis, or by a suitable zero padding. 
In our CFD application, however, axial stretching is not a viable solution, as the image information would be altered. 
Zero padding has been tried, but with largely different sizes at play there are instances where an image would predominantly consist of zeros, which causes the so called problem of dead neurons \citep{lu-etal-2020}, and eventually deteriorates the learning abilities of the NN.
Therefore, in our procedure images are split into groups of the same size, and loaded at training time in batches, where each batch only contains images of the same group, hence of the same size. Different batches can of course contain images of different sizes. 

\section{The turbulence models}
\label{sec:models}

ATM can be implemented in several forms; in principle, any set of turbulence models could be blended together. The specific form impacts the complexity of the integration of ATM with the RANS solver.
For example, with the classic zonal modelling \cite{avva-kline-ferziger-1988} blended six different implementations of the $k - \epsilon$ model. 

In this preliminary work, in agreement with the minimalist dictionary that only contains three structural flow zones, only three RANS turbulence models are employed.
The considered models are the $k-\epsilon$ model \citep{launder-spalding-1974}, the $k-\omega$ model \citep{wilcox-1988}, and an extension of the $k-\omega$ model which features an additional rotation/curvature correction \citep{hellsten-1998}. 
Overall, our formulation resembles the $k-\omega$ SST model, introduced by \cite{menter-1993}, in which a blending function $F_1$ is used to locally switch between $k-\epsilon$ and $k-\omega$, with the major difference that $F_1$ in the conventional SST formulation is empirically determined based on various flow quantities and on the distance from the wall, whereas here $F_1$ is determined at runtime from the segmentation of the flow field. 

The PDEs describing an unified formulation of the three models are the following two model equations for the turbulent kinetic energy $k$ and for the turbulent frequency $\omega$:

\begin{equation}
\frac{\partial (\rho k)}{\partial t} + \frac{\partial (\rho U_j k)}{\partial x_j}= P - \beta^* \rho \omega k + \frac{\partial}{\partial x_j} \left[ (\mu + \sigma_k \mu_t) \frac{\partial k}{\partial x_j} \right] ,
\end{equation}
\begin{multline}
\frac{\partial (\rho \omega)}{\partial t} + \frac{\partial (\rho U_j \omega)}{\partial x_j} = \frac{\gamma}{\nu_t} P - (F_4 F_r + 1 - F_r) \beta \rho \omega^2 + \\
+ \frac{\partial}{\partial x_j} \left[ (\mu + \sigma_\omega \mu_t) \frac{\partial \omega}{\partial x_j} \right] + 2 (1 - F_1) \frac{\rho \sigma_\omega}{\omega} \frac{\partial k}{\partial x_j} \frac{\partial \omega}{\partial x_j}
\label{eq:omegaeqn}
\end{multline}
in which $U_j$ indicates the $j$-th mean velocity component, repeated indices imply summation, $P = \tau_{ij} \partial U_i / \partial x_j$ is the production of turbulent kinetic energy, $\tau_{ij} = \mu_t \left(2S_{ij} - 2/3 \partial u_k / \partial x_k \delta_{ij} \right) - 2/3 \rho k \delta_{ij}$ are the Reynolds stresses expressed through the Bousinnesq closure, and $S_{ij}$ is the symmetric part of the mean velocity gradient tensor. 
The coefficient $\beta^*$ has the same value in all models, while the coefficients $\sigma_k, \gamma, \beta$ and $\sigma_\omega$ are locally a blend, through the function $F_1$, of the values employed in pure $k-\epsilon$ and $k-\omega$. Lastly, the turbulent dynamic viscosity is computed via the algebraic relation $\mu_t = \rho k / \omega$.
The cross-diffusion term between $k$ and $\omega$, i.e. the last term in Eq.\eqref{eq:omegaeqn}, is missing in the $k-\omega$ model; the factor $1-F_1$ is responsible for the model being intermediate between pure $k-\epsilon$ (when $F_1=0$) and pure $k-\omega$ (when $F_1=1$).
In the original SST formulation, the $F_1$ function is empirical, and expressed as:
\begin{equation}
F_1 = \tanh \left( arg_1^4 \right) ,
\end{equation}
\begin{equation}
arg_1 = \min \left[ \max \left( \frac{\sqrt{k}}{\beta^*\omega d},
   \frac{500 \nu}{d^2 \omega} \right) , \frac{4 \rho \sigma_{\omega 2} k}{CD_{k \omega} d^2} \right] ,
\end{equation}
\begin{equation}
CD_{k \omega} = \max \left(2 \rho \sigma_{\omega 2} \frac{1}{\omega}
   \frac{\partial k}{\partial x_j} \frac{\partial \omega}{\partial x_j}, 10^{-20} \right) ,
\end{equation}
where $\rho$ is the fluid density, $\nu$ is the molecular kinematic viscosity, $d$ is the distance of a point from the nearest wall, and $\beta^*$ and $\sigma_{\omega 2}$ are two model coefficients. 

The curvature correction is applied via the multiplicative factor $F_4$ in front of the dissipation term $\beta\rho\omega^2$ of the $\omega$ equation \eqref{eq:omegaeqn}. In our formulation the binary function $F_r$ controls where the correction is on ($F_r = 1$) or off ($F_r =0$), whereas in the original formulation \citep{hellsten-1998} no $F_r$ is present as the curvature correction is always on. The factor $F_4$ is defined as follows:
\begin{equation}
  F_4 = \frac{1}{1+C_{RC} R_i} ,
\end{equation}
\begin{equation}
  R_i = \frac{\Omega}{S} \left( \frac{\Omega}{S} - 1 \right) ,
\end{equation}
where $S$ and $\Omega$ are defined as $S=\sqrt{2S_{ij}S_{ij}}$ and $\Omega=\sqrt{2 \Omega_{ij} \Omega_{ij}}$. The tensor $\Omega_{ij}$ is the anti-symmetric part of the mean velocity gradient tensor; the coefficient $C_{RC}$ is set to the value $1.4$ following \cite{mani-ladd-bower-2004}.

We assign $F_1=1$ (therefore pure $k-\omega$ model) to regions with labels {\em R} and {\em BL}, and $F_1=0$ (therefore pure $k-\epsilon$) elsewhere, i.e. regions with labels {\em SL} and $\circ$. Moreover, we set $F_r=1$ (active curvature correction) in regions with label {\em R}.
This choice is motivated by the evidence that the $k-\omega$ model is proved to work better than $k-\epsilon$ for adverse pressure gradient and boundary layer flows \citep{pope-2000}, while $k-\epsilon$ works better in shear layers and has less trouble with the free-stream boundary condition \citep{menter-1993}. 
Improving ATM above this rather simplistic starting point is where the expert contribution from the RANS modelling community is deemed necessary.

The zonal model described above is implemented within the open-source C++ library OpenFoam \citep{weller-etal-1998}, where the new turbulence model is compiled in the library. 
To combine the output of the segmentation with the new model, the segmentation and the calculation of $F_1$ and $F_r$ are embedded into the iterative computation of the CFD solution. The simulation starts with $F_1=1$ and $F_r=0$ everywhere. Segmentation (and the subsequent rearrangement of the spatial distribution of $F_1$ and $F_r$) is carried out a first time whenever the largest residual reaches $10^{-3}$, and then repeated whenever a further drop by one order of magnitude is observed. 
This workflow provides results that are independent from an arbitrary choice of the initial condition of $F_1$ and $F_r$, as will be shown in \S\ref{sec:verification}.

\section{Performance of the automatic segmentation}
\label{sec:results-segmentation}

Before demonstrating the entire ATM procedure, we discuss here the performance of the NN for semantic segmentation. 
The NN is first applied in \S\ref{sec:segmentation-laminar} to the validation test set, which consists of manually annotated laminar flow cases; in a second step, in \S\ref{sec:segmentation-turbulent} the NN is used to segment turbulent flows computed with conventional RANS modeling.

\subsection{Laminar flows}
\label{sec:segmentation-laminar}

\begin{figure}
\centering
\includegraphics[width=0.3\textwidth]{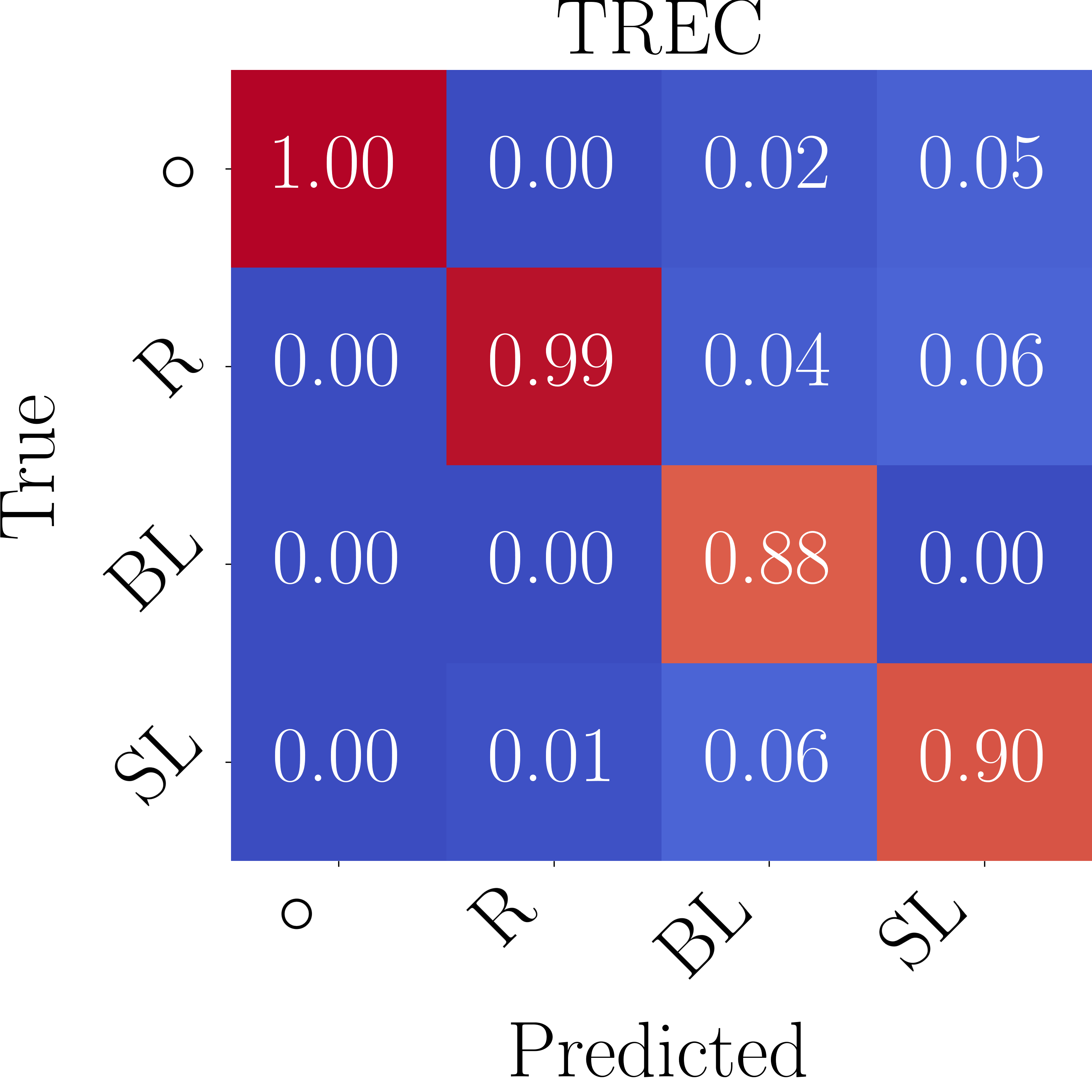}
\includegraphics[width=0.3\textwidth]{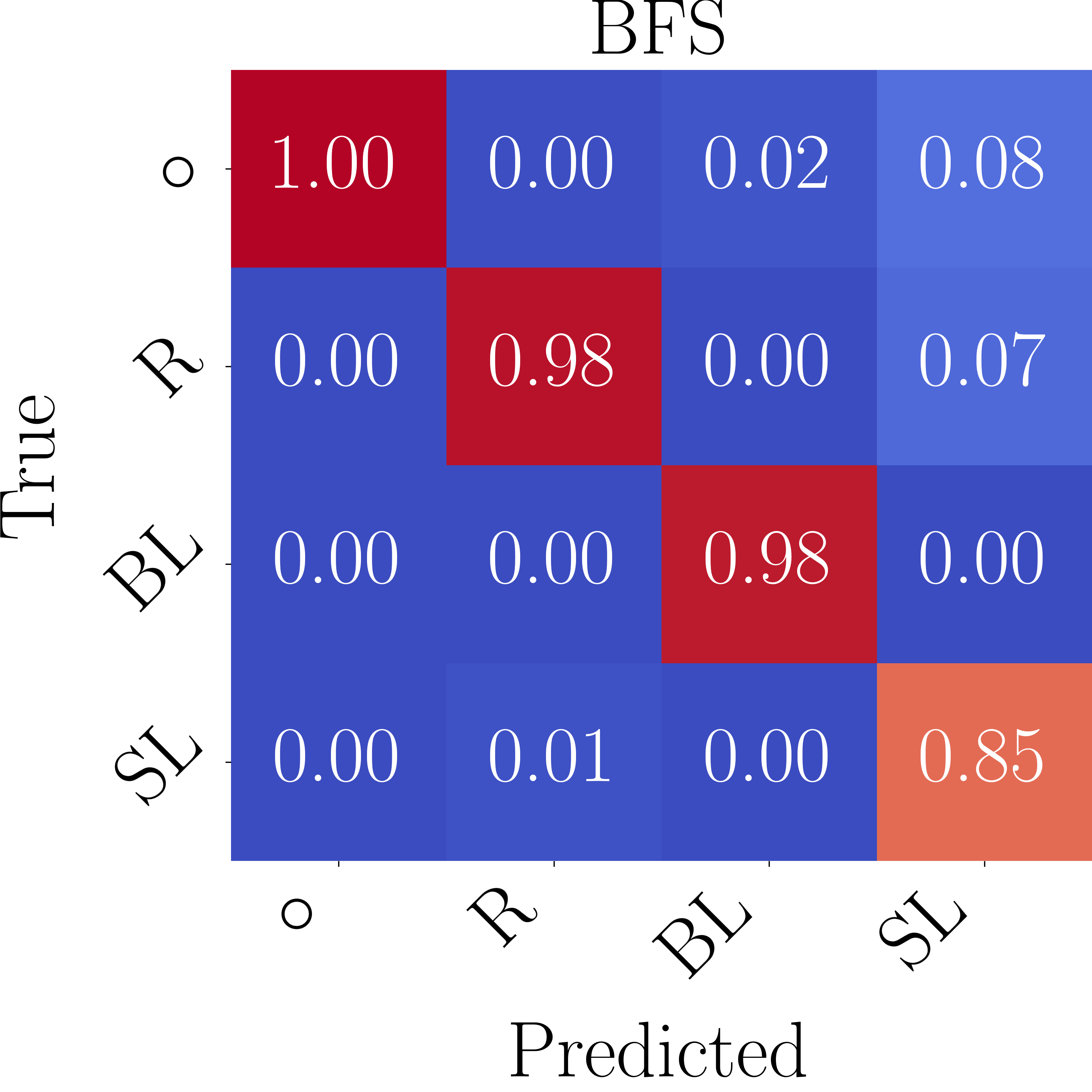}
\includegraphics[width=0.3\textwidth]{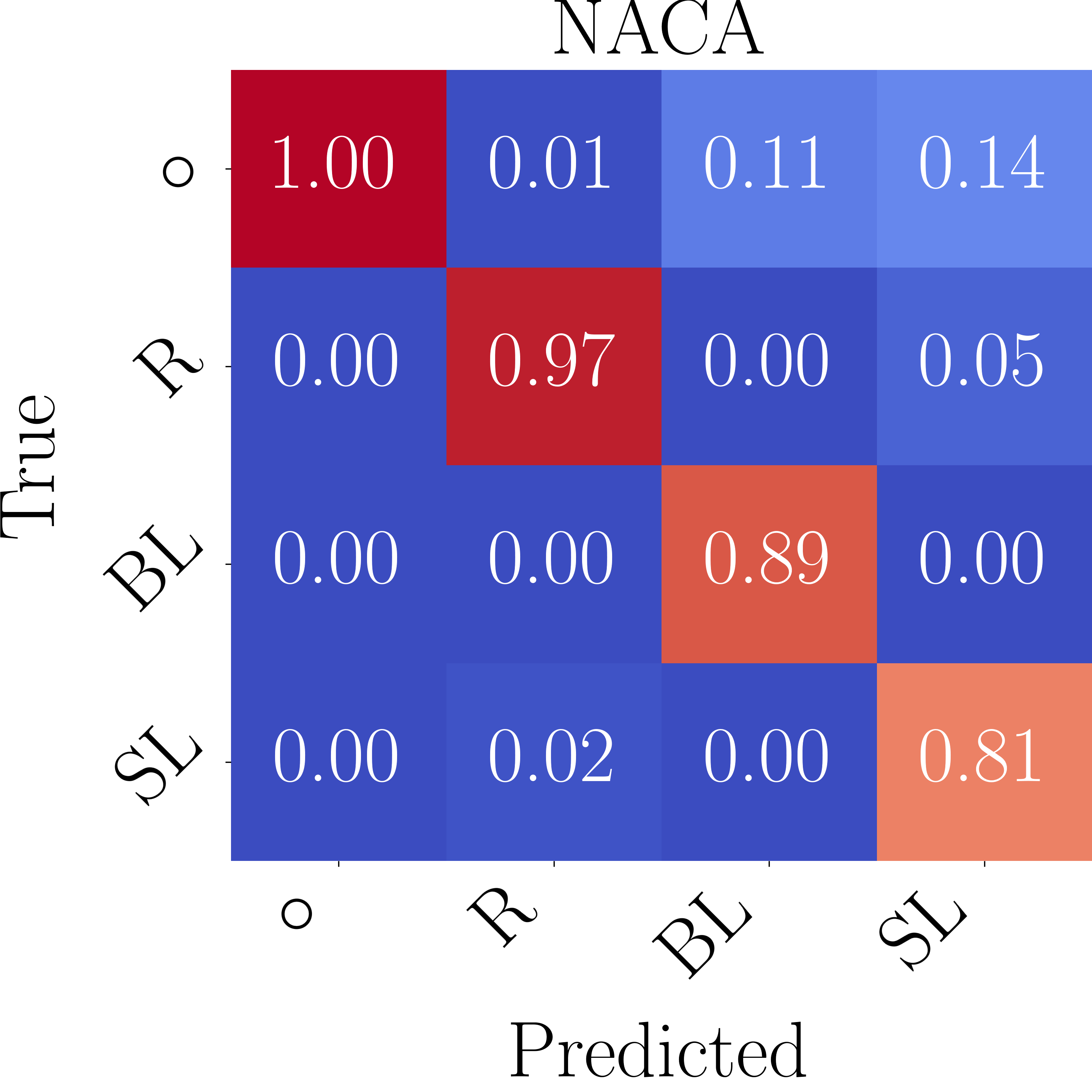}
\caption{Confusion matrices for the TREC (left), BFS (center) and NACA (right) flow cases. The diagonal elements represent the percentage of true positive pixels. Off-diagonal elements on a row are false negatives, and on a column are false positives. The color scale is from blue (zero) to red (one).}
\label{fig:confusionmatrices}
\end{figure}

Using the training dataset described in \S\ref{sec:dataset}, which includes five different flows, five distinct NN are trained, by removing one flow class at a time, and subsequently tested on the unseen class against the manual annotations. 
Since ground-truth annotations are available, the ability of the NN to segment the flow field correctly can be quantified via confusion (or error) matrices \citep{tharwat-2020}. 
A confusion matrix is a square matrix, whose rank equals the number of labels: each row of the matrix corresponds to the true class, while each column represents the predicted class. Hence, a confusion matrix displays correct classifications (true positives) on the diagonal elements, whereas false-positive and false-negative predictions for each label appear in the off-diagonal elements of the columns and row, respectively, of each label.
Three of these confusion matrices, normalized by column, are plotted in figure \ref{fig:confusionmatrices} for classes TREC, BFS and NACA. For brevity, the remaining two matrices are not reported; their results are comparable to or better than the ones discussed here. Also, the row and column pertaining to the solid body label {\em X}, whose prediction is trivial, are removed.

All matrices are strongly diagonally dominant, which in general indicates good predictions.
Most errors, particularly for the NACA class, derive from pixels of the featureless flow class $\circ$ being assigned the label {\em BL} or {\em SL} (third and fourth element of the first row). 
A limited degree of confusion between one of the flow features and the background is acceptable, as it only affects the borders of the segmented region, whose manual annotation is already arbitrary to some extent, and thus only results in a slight shift of the boundary between the background model and the zone-specific model. 
Moreover, once the dictionary will grow to include more words, the number of pixels with featureless background will diminish, rendering this problem progressively less important.
Little confusion is present where some {\em R} pixels are classified as {\em SL} (as visible from elements (2,4) of the matrices). 
This confusion is somehow reasonable as label {\em SL} starts at the edge of label {\em R} (as visible from figure \ref{fig:dataset}). In perspective, it is imperative for the network to correctly discriminate flow features, keeping this type of confusion to a minimum.

\begin{table}
\centering
\begin{tabular}{l|ccccc}
           & REC   & TREC   & BFS   & DBFS  & NACA     \\ 
\hline
{\em R}    & 0.97  & 0.96   & 0.96  & 0.98  & 0.96     \\ 
{\em BL}   & 0.96  & 0.87   & 0.97  & 0.96  & 0.86     \\   
{\em SL}   & 0.89  & 0.80   & 0.78  & 0.84  & 0.72     \\
\end{tabular}
\caption{$IoU$ score for the labels {\em R}, {\em BL} and {\em SL}.}
\label{table:iouresults}
\end{table}

To provide a compact representation of the segmentation accuracy, several metrics can be calculated from the (non-normalized) confusion matrices.
Here we opt for the $IoU$ score represented by equation \eqref{eq:iou}, which is often used to quantify the quality of segmentations \citep{garcia-etal-2017}. 
Table \ref{table:iouresults} reports, for the 5 NNs, the $IoU$ score for the segmentation of the three labels {\em BL}, {\em R} and {\em SL}. 
The overall performance is more than satisfactory: the network generalizes well on the flow case excluded from training, and consistently achieves high scores.
It is noteworthy that good scores are obtained for the NACA class too. This class is quite different from the remaining four used in training, and is the only occurrence of an external flow around a streamlined body. Therefore, the good performance of the NN for the NACA class is an indirect indication of the network ability to generalize.

\subsection{Turbulent flows}
\label{sec:segmentation-turbulent}

We now proceed to applying the segmentation network to the output of RANS simulations, to confirm with two examples that the network trained on a laminar dataset performs well on turbulent flow fields.

\begin{figure}
\centering
\includegraphics[width=0.8\textwidth]{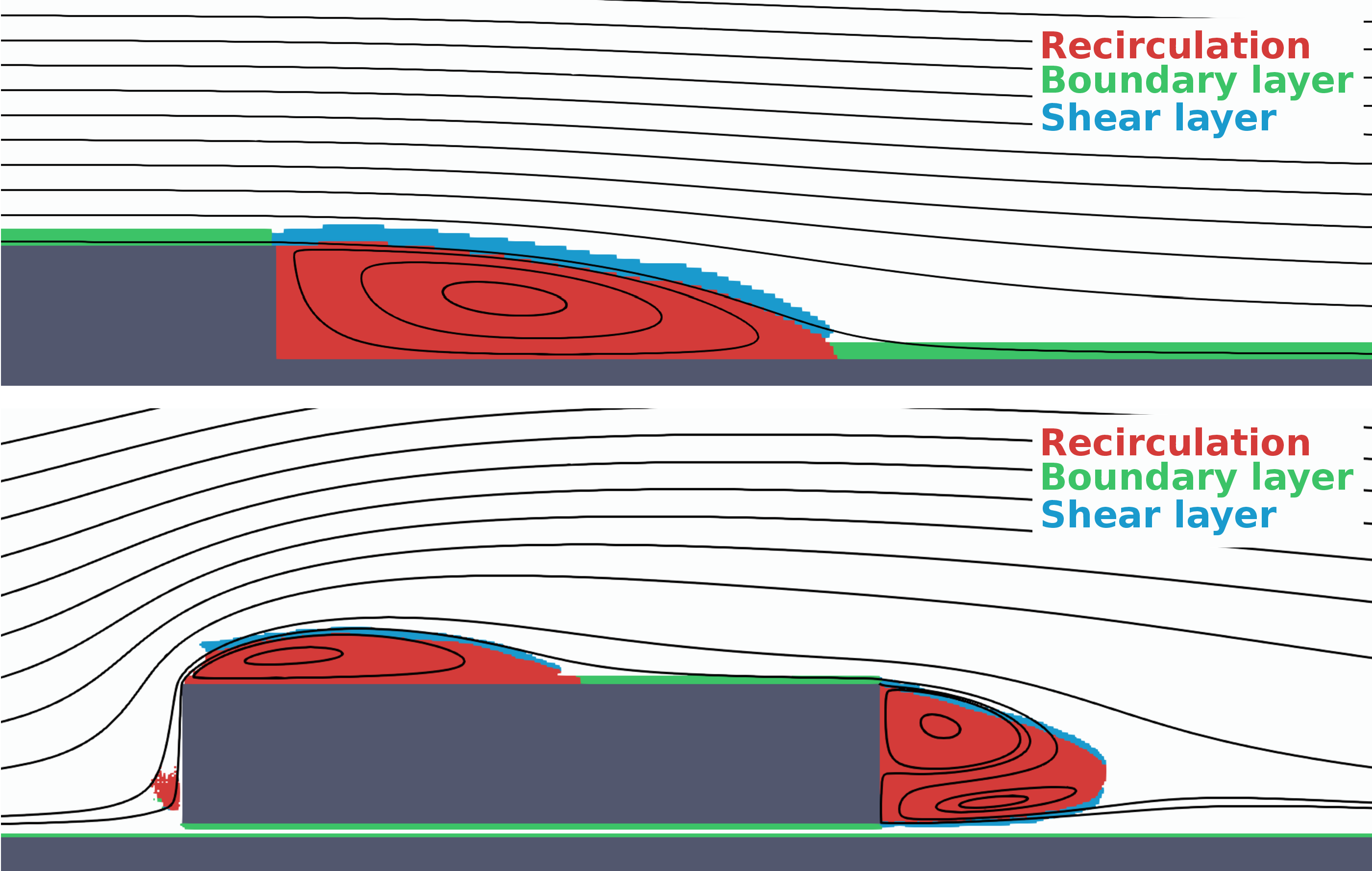}
\caption{Segmentation of RANS-computed flow fields: open channel with backward-facing step (top) and rectangular body near a solid wall (bottom). The RANS solution is represented via its streamlines. The NN segmentation is represented in color: red indicates label {\em R}, green is label {\em BL}, and light blue is label {\em SL}.}
\label{fig:net-turbulent}
\end{figure}

The first example is an open channel flow where the wall presents a backward-facing step. This configuration, that will be also used later on to evaluate the performance of the ATM, is considered here to the sole purpose of verifying the segmentation. 
It represents a flow that is already contained in the training dataset, albeit in its laminar version.
The RANS simulation employs here a standard RANS model (namely, the $k-\epsilon$ model); geometry and Reynolds number mimic those of the DNS study carried out by \cite{le-moin-kim-1997}. The value of Reynolds number, based on the step height, is $Re=5100$. 
Figure \ref{fig:net-turbulent} (top) shows the RANS solution, expressed in terms of streamlines, with the computed segmentation superimposed in color, according to the convention introduced in figure \ref{fig:dataset}. 
The agreement between the segmentation and the RANS solution is extremely satisfactory.

Figure \ref{fig:net-turbulent} (bottom) is for a further test case, that is not contained in the training dataset: a rectangle with $\AR = 5$ is placed near a solid wall, with $Re=3000$ based on the length of the vertical side.
Although apparently not too dissimilar from the previous example, this second flow contains a number of physical features that are absent in the training set: in particular, the non-symmetric recirculation and the interaction with the wall.
It can be visually appreciated that, as expected, predictions are not as accurate as the previous ones, but remain very satisfactory: every flow class, even though not perfectly segmented, is identified very well. In particular, the non-symmetric recirculation at the rear, and the double boundary layer in the bottom gap are perfectly identified. 
Only a very small patch in front of the rectangle is misclassified as recirculation: here the NN is confused by the large vertical velocity, caused by the flow impingement on the anterior part of the body. Such error is clearly related to the lack of the corresponding feature (impingement) in the training dataset; indeed, creating a general and comprehensive dataset is key for training robust NNs.

\section{Predictive capabilities of ATM}
\label{sec:results-model}

Two examples are used in this section to illustrate the performance of the automated turbulence modeling procedure. 
The examples feature a two-dimensional mean flow, as requested by the present, minimalist implementation, and are geometrically simple, so that complexities related to discretization and meshing are avoided. 
Moreover, they have been designed to be meaningful with respect to the labels the NN has been trained to identify; in both cases, high-fidelity (DNS or experimental) data are available to gauge the quality of the solution.

The first example concerns a simple backward-facing step, that is often employed as a test case to study the performance of turbulence models in presence of flow separation, since the step corner removes the uncertainty related to the position of the separation point. 
The flow is an easy segmentation task, since similar (laminar) flows are contained in the training set of the NN. 
The second example is another standard test case, a wall-mounted hump, where recirculation obviously plays a major role. This type of flow was not considered in the training set, and is therefore meaningful for evaluating the entire procedure.

In both examples, our implementation of ATM will be compared against the two baseline models $k-\epsilon$ and $k-\omega$, as well as the gold standard represented by the $k-\omega$ SST model.

\subsection{Example: turbulent flow over a backward-facing step}
\label{sec:step}

\subsubsection{Description}

The first application example consists in an open channel flow bounded by a wall with a backward-facing step. The reference is provided by the direct numerical simulation (DNS) results of \cite{le-moin-kim-1997}. 
Since this flow is among those used for the training of the segmentation network, it can be safely assumed that the NN works well. This has been already found to be the case in \S\ref{sec:segmentation-turbulent} (albeit on a mean flow computed with standard turbulence modelling), and will not be considered further.

\cite{le-moin-kim-1997} solved by DNS the incompressible Navier--Stokes equations. The Reynolds number of their study is $Re_h = U_0 h / \nu = 5100$, based on the step height $h$ and the maximum $U_0$ of the mean velocity profile at the inlet. 
The computational domain has a total streamwise length of $30h$, including a length of $10h$ upstream the step, that is located at $x/h = 0$. The vertical domain height after the step is $6h$. 
The major feature of the flow is a large recirculating region with length of $6.28 h$; the end of the recirculating region was determined in the original study by locating the point of zero wall-shear stress. 
A shorter secondary recirculation bubble develops near the step corner, with a length of $1.76 h$, and extends for $0.8 h$ in the vertical direction, when measured along the vertical wall of the step.

DNS results are compared to RANS results computed at the same $Re$, and using the same boundary conditions: no-slip and no-penetration conditions at the solid wall, and a free-slip condition at the upper boundary. As for the outlet, DNS employs a time-dependent outflow condition $\partial u_i / \partial t + U_c \partial u_i / \partial x =0$ for each velocity component $u_i$, where $U_c$ is the (constant) mean exit velocity. This condition is translated into a zero gradient condition for the RANS simulation. The inlet in RANS is uniform for simplicity, and it is verified that, $3h$ upstream of the step, the velocity profile is developed and its maximum velocity yields the correct value used in the definition of $Re$.
An identical structured grid is used in all the RANS simulations, for which grid sensitivity is checked first to verify its suitability. 

\subsubsection{Comparison with conventional RANS models}

The performance of a turbulence model in this flow is often assessed \citep[see for example][]{lasher-taulbee-1992, akselvoll-moin-1993, bredberg-peng-davidson-2002} by examining the streamwise evolution of the skin-friction coefficient and the values of the reattachment lengths. The former, considered only downstream of the step,  is a dimensionless measure of the wall-shear stress $\tau_w$, and is defined as
\[
C_f = \frac{2 \tau_w}{\rho U_0^2} .
\]

\begin{table}
\centering
\begin{tabular}{ccccc}     
& $k-\epsilon$  & $k-\omega$   & SST   & ATM \\ 
\hline
$\%\Delta X^{(1)}_r$ & $-21.0$ & $+14.6$ & $+14.0$ & $+3.3$ \\ 
$\%\Delta X^{(2)}_r$ & $-80.1$ & $ -1.5$ & $ -3.2$ & $-4.6$ \\
$\%\Delta Y^{(2)}_r$ & $-62.5$ & $ -4.1$ & $ -3.0$ & $-5.1$ \\
\end{tabular}
\caption{Length $X_r$ and width $Y_r$ of the primary and secondary recirculating bubbles for the backward-facing step flow: percentage difference between RANS and DNS.}
\label{table:reclengths}
\end{table}

The streamwise length $X_r^{(1)}$ of the primary recirculation bubble (whose width equals the step height), and the length and width $X_r^{(2)}$ and $Y_r^{(2)}$ of the secondary recirculation bubble near the corner are reported in table \ref{table:reclengths} in terms of percentage differences between RANS predictions and DNS information. 
The $k-\epsilon$ model severely underpredicts the length of the primary recirculation by more than 20\%. This model in fact is known to have issues at properly representing the normal Reynolds stress differences that play an important role in adverse pressure gradient flows \citep{speziale-ngo-1988}. 
$k-\omega$ and the SST variant both overpredict the same length, by approximately 14\%. ATM provides the best result here, with a minor overprediction of 3\% only. 
The horizontal size of the secondary recirculating regions is well predicted by all models, with the notable exception of $k-\epsilon$.
The vertical extent of the secondary recirculation, that is not constrained by geometry, is once again quite far off for $k-\epsilon$, which predicts a recirculation that is 80\% shorter and 60\% smaller in the vertical direction. The other three models are again comparable and within 5\%.

\begin{figure}
\centering
\includegraphics[width=0.8\textwidth]{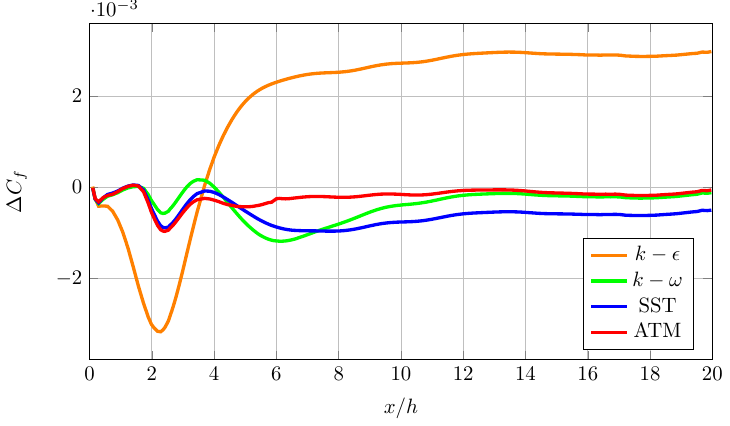}
\caption{Evolution of the error in the skin-friction coefficient along the wall past the step: difference $\Delta C_f$ between the RANS predictions and the DNS measurement.} 
\label{fig:deltacf}
\end{figure}

A more detailed view of the model performance can be obtained by looking at the entire evolution of $C_f$ along the wall, thus going beyond its zero crossings which determine the size of the recirculating regions. Figure \ref{fig:deltacf} plots the streamwise evolution after the step of the difference $\Delta C_f$ between the RANS-predicted friction coefficient and the one computed by DNS and assumed as reference.
The $k-\epsilon$ model, besides failing significantly at retrieving the correct recirculation lengths, provides a poor prediction for $C_f$ everywhere in the domain, including the downstream recovery zone, where the error is of the order of 100\%.
The performance of ATM, especially within the primary recirculating bubble, is more or less on par with the best models, and provides a balanced prediction across the entire extent of the channel, distinguishing itself for accuracy in the region $5 < x/h < 12$.

\begin{figure}
\centering
\includegraphics[width=\textwidth]{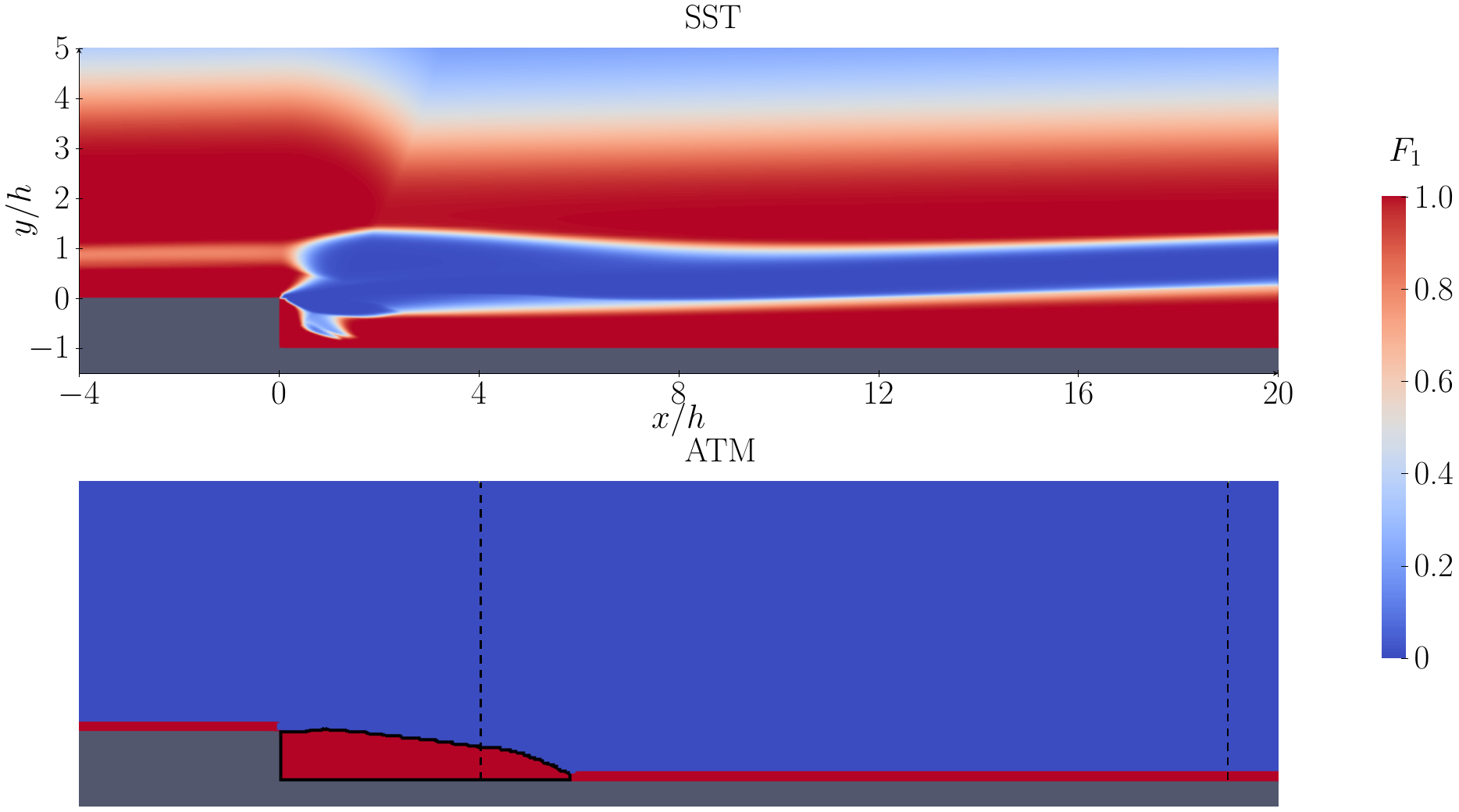}
\caption{Comparison between the empirical blending function $F_1$ of the SST model (above) and the one implicitly computed with ATM (below). $F_1=1$ corresponds to pure $k-\omega$, and $F_1=0$ corresponds to pure $k-\epsilon$. The black contour on the lower image draws the boundary of the region where $F_r=1$ and the curvature correction is on. The two vertical dashed lines mark the streamwise locations $x/h=4$ and $x/h=19$, used later in figure \ref{fig:profiles}.}
\label{fig:f1}
\end{figure}

Figure \ref{fig:f1} compares the empirical blending function $F_1$ computed by the SST model against the one defined implicitly with ATM. SST mandates the use of $k-\omega$ across most of the channel, but for a rather well defined region that starts at the step corner and reaches the outflow boundary while slowly moving towards the centerline. Shape and extension of this region have no clear physical interpretation. Near the free-slip boundary, $F_1$ is such that the model inclines more towards $k-\omega$ near the inflow, and is somewhat uncertain between the two constitutive models for the downstream part. The inflow is mostly $k-\omega$, and the outflow intertwines both models.
ATM, instead, chooses between the two models in a more ordered way. It limits the use of $k-\omega$ to the near-wall region beginning-to-end, and to the primary recirculation zone, inside which the curvature correction is switched on.
Just outside the recirculation, where a shear layer exists, ATM switches to $k-\epsilon$. The bulk flow and the free-slip boundary are assigned to $k-\epsilon$ as the model of choice for the featureless background. 
Since $k-\epsilon$ better captures the physics of a shear layer \citep{menter-1993}, and is capable of a more accurate representation of turbulent shear stresses, ATM has an edge here and achieves better performance compared to SST.

\begin{figure}
\centering
\includegraphics[width=\textwidth]{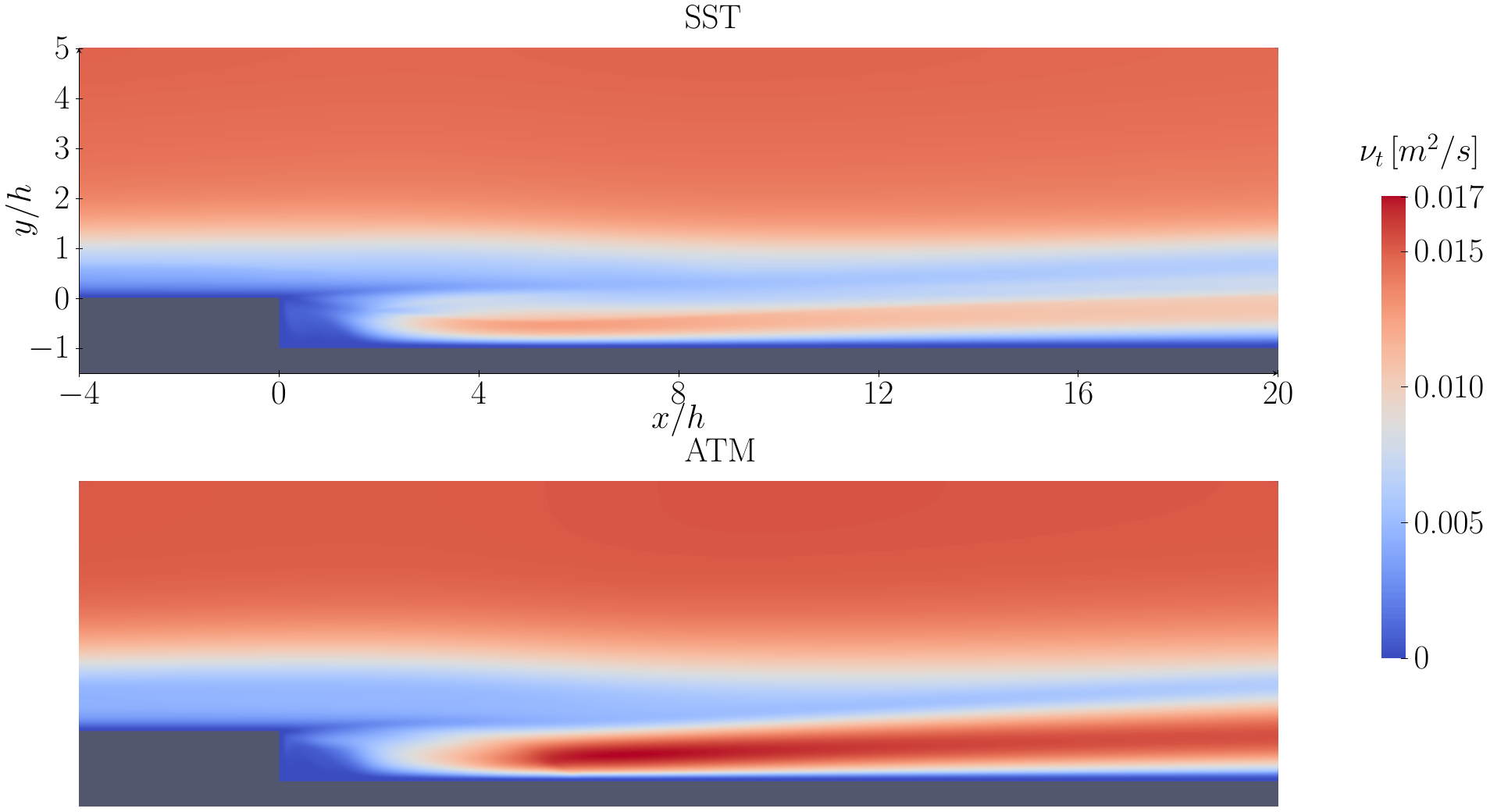}
\caption{Comparison between the field of the eddy viscosity $\nu_t$ of the SST model (above) and the one computed within ATM (below).}
\label{fig:nut}
\end{figure}

Figure \ref{fig:nut} compares the field of turbulent viscosity $\nu_t$ as computed by the SST model and by ATM. The spatial distribution of $\nu_t$ and its amplitude are reasonably similar, but not identical: ATM tends to prescribe higher turbulent viscosities downstream of the main recirculation.
The most important observation though is that, at least in this specific implementation of ATM, eddy-viscosity-based models are combined together by ATM with a relatively sharp transition, but this does not cause any jump in the field of $\nu_t$.

\begin{figure}
\centering
\includegraphics[width=\textwidth]{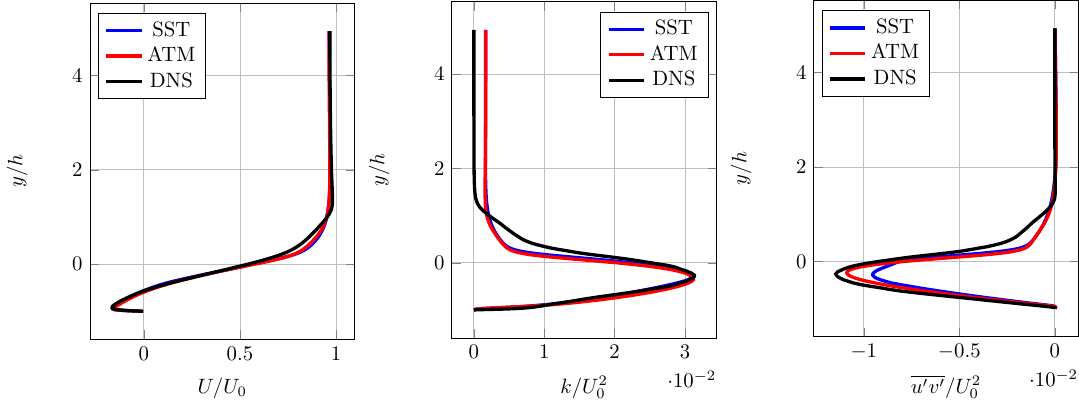}
\includegraphics[width=\textwidth]{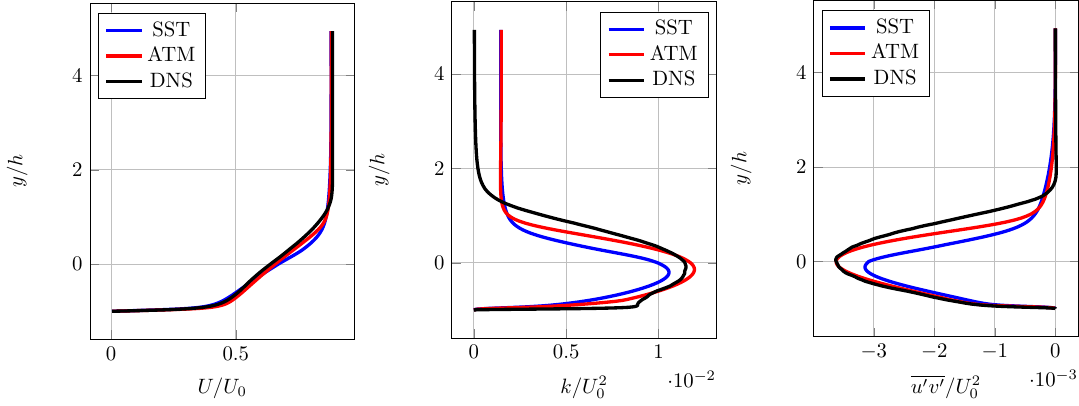}
\caption{Wall-normal profiles of the mean horizontal velocity (left), turbulent kinetic energy (center) and turbulent shear stress (right) inside the primary recirculation bubble at $x/h=4$ (top) and after reattachment at $x/h=19$ (bottom): comparison between ATM and SST against DNS data.}
\label{fig:profiles}
\end{figure}

To observe the solution quality inside the flow field, figure \ref{fig:profiles} compares with DNS data the wall-normal profiles for the mean velocity (left), the turbulent kinetic energy (center) and the turbulent shear stress (right), at the two streamwise location $x/h=4$ (inside the primary recirculation) and $x/h=19$ (well after reattachment) shown in figure \ref{fig:f1}. 
At $x/h=4$ the mean velocity and the turbulent kinetic energy profiles are pretty similar between ATM and SST, yet the ATM prediction of the turbulent stress is closer to the one of the DNS. As explained by \cite{rumsey-etal-2004}, a correct prediction of $\overline{u'v'}$ is important for getting a correct size of the recirculating bubble. 
At the downstream station $x/h=19$, the mean velocity profiles are quite similar, but the SST curve shows a small, non-physical bump at $y/h \approx 0.5$, in correspondence to the sudden change (see figure \ref{fig:f1}) of the blending function from $F_1=1$ to $F_1=0$. The turbulent kinetic energy profile shows the maximum predicted by ATM is very near to the DNS maximum, and overall is much closer to the DNS profile than the SST profile. At both streamwise stations, ATM is also more accurate at the prediction of the turbulent shear stress profile.

\subsection{Example: turbulent flow over a wall-mounted hump}
\label{sec:hump}

\subsubsection{Description}

Another popular test case for turbulence models is the 2D NASA wall-mounted hump (WMH in the following). The hump consists of a relatively long forebody and a relatively short concave ramp in the aft.
The flow over the WMH was studied in a series of experimental works by \cite{greenblatt-etal-2006b, greenblatt-etal-2006} and \cite{naughton-viken-greenblatt-2006}. 
Experimental data extracted from these experiments are often used as a reference for evaluating the performance of RANS models \citep{rumsey-etal-2004}, and will be also used here as the baseline solution to compare with.
The WMH test case probes the ability of turbulence models to correctly predict separation, reattachment and recovery induced by an adverse pressure gradient on smoothly contoured bodies.
The typical model tends to underestimate the turbulent shear stress within the separated shear layer, thus overpredicting the length of the separated region \citep{rumsey-etal-2004}. 

The Reynolds number in the experiments is set to $Re_c = U_0 c / \nu = 9.29 \times 10^5$ based on the chord $c$ of the hump, and on the free-stream velocity $U_0$ of the incoming boundary layer. The Reynolds number is rather large and implies a fully turbulent flow field.
The separation point is located at $x/c = 0.66$, and the flow reattaches at $x/c =1.11$.
The experiments provide profiles for mean velocity and turbulent stresses, measured at different stations along the streamwise direction, as well as for the evolution of the skin-friction coefficient along the entire lower wall.

The RANS simulations are configured to replicate the experimental set up. The WMH geometry and the computational grid are taken without modifications from the NASA Turbulence Modeling Resources (TMR) website (https://turbmodels.larc.nasa.gov/index.html); in particular their second-finest grid is chosen, consisting of 817 cells in the streamwise direction and 217 cells in the vertical directions. 
The simulation domain extends for $-6.39 < x/c < 4$ in the horizontal streamwise direction, with the hump located between $0 < x/c < 1$, and for $0 < y/c < 0.128$ in the vertical direction. 
At the wall the no-slip condition is enforced. The NASA geometry also includes a contoured upper boundary, to account for side-plate blockage: here the inviscid no-penetration condition is applied, according to the prescription of the TMR website. An uniform velocity profile is imposed at the inlet to retrieve the correct value of $U_0$ upstream the hump. 

\subsubsection{Comparison with conventional RANS models}

\begin{figure}
\centering
\includegraphics[width=\textwidth]{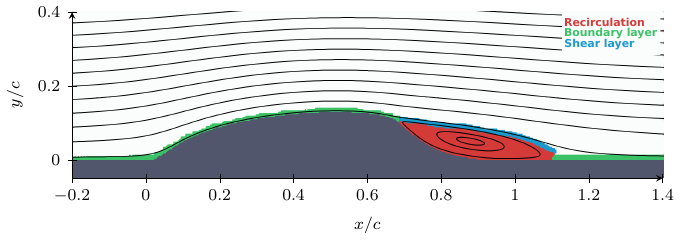}
\caption{Segmentation of the WMH flow field. The RANS
solution is represented via its streamlines, and colours correspond to labels: red is {\em R}, green is {\em BL} and light blue is {\em SL}.}
\label{fig:wmh-segmentation}
\end{figure}

First, figure \ref{fig:wmh-segmentation} illustrates the segmentation of the flow field obtained, at convergence, by the NN. Although the network has been trained on laminar flows only, and has never seen this specific flow, the segmentation appears to be entirely accurate, with all the flow regions properly identified. 

\begin{table}
\centering
\begin{tabular}{ccccc}
    & $k-\epsilon$   & $k-\omega$    & SST    & ATM    \\ 
\hline
$\%\Delta X_r$   & $+8.8$    & $+7.3$    & $+11.3$   & $+1.9$   \\ 
\end{tabular}
\caption{Length of the recirculating region: percentage difference $\%\Delta X_r$ between RANS and experimental information.}
\label{tab:wmhrec}
\end{table}

The assessment of ATM considers the length of the recirculating region, and the evolution of the skin-friction coefficient along the wall.
The position of the reattachment point as predicted by the different turbulence models is presented in table \ref{tab:wmhrec} in terms of percentage difference with respect to the experimental value, taken as reference. 
As in the previous example, ATM predicts an essentially correct reattachment length, within 2\% of experimental information, and performs significantly better than its constituent models, including SST, which has an 11\% error.

\begin{figure}
\includegraphics[width=\textwidth]{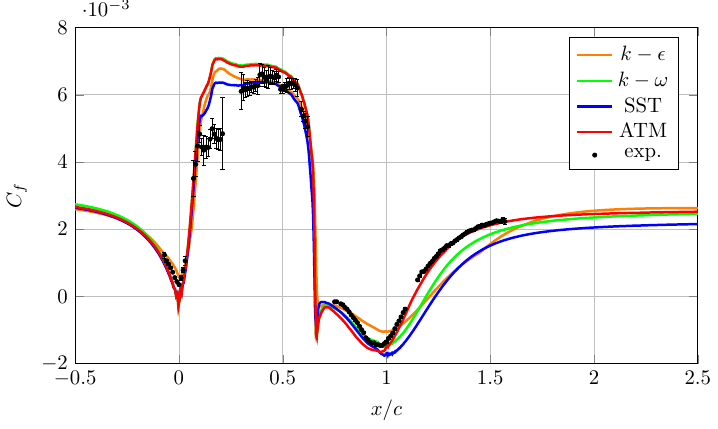}
\caption{Evolution of the friction coefficient $C_f$ at the wall: comparison between ATM and other turbulence models against experimental data; the error bars indicate experimental uncertainty. The hump is located at $0 < x/c < 1$.}
\label{fig:wmhcf}
\end{figure}

Figure \ref{fig:wmhcf} plots the evolution of the friction coefficient $C_f$ along the lower wall, and compares it with experimental information. 
The friction coefficient above and after the hump presents three distinct regions. In a first region, for $0 < x/c < 0.6$, that extends from the beginning of the hump to the separation point, $C_f$ increases sharply, then saturates on top of the hump, and then abruptly decreases towards zero; a second region, for $0.6 < x/c < 1.1$, shows a negative $C_f$ and indicates the extent of the recirculation; in a third region for $x/c > 1.1$, $C_f$ assumes positive values again, and recovers towards a constant value well after the reattachment point. 

All the considered models capture the qualitative evolution of the skin-friction coefficient. 
In the first region, the experimental uncertainty is large, and results have therefore to be interpreted with caution, ATM and $k-\omega$ overlap, and overestimate the maximum level of friction above the hump tip. 
SST and, to a lesser extent, $k-\epsilon$ do a bit better.
In this zone the boundary layer is strongly accelerated, but wall flows with favourable pressure gradients are not included in the present dictionary of flow zones. Hence, it is reasonable to expect that including the accelerated boundary layer as an additional flow zone would improve the predictive capabilities of ATM in this region.
In the second region, where recirculation takes place, ATM and $k-\omega$ capture correctly intensity and position of the recirculating bubble, while $k-\epsilon$ underestimates the intensity, and SST overpredicts it. 
Lastly, in the third region, the prediction by ATM is best, and achieves a very close agreement with experimental data, whereas the other models, and in particular SST, seem to recover the flat-plate level of friction only after an excessively long transient. 

\begin{figure}
\centering
\includegraphics[width=\textwidth]{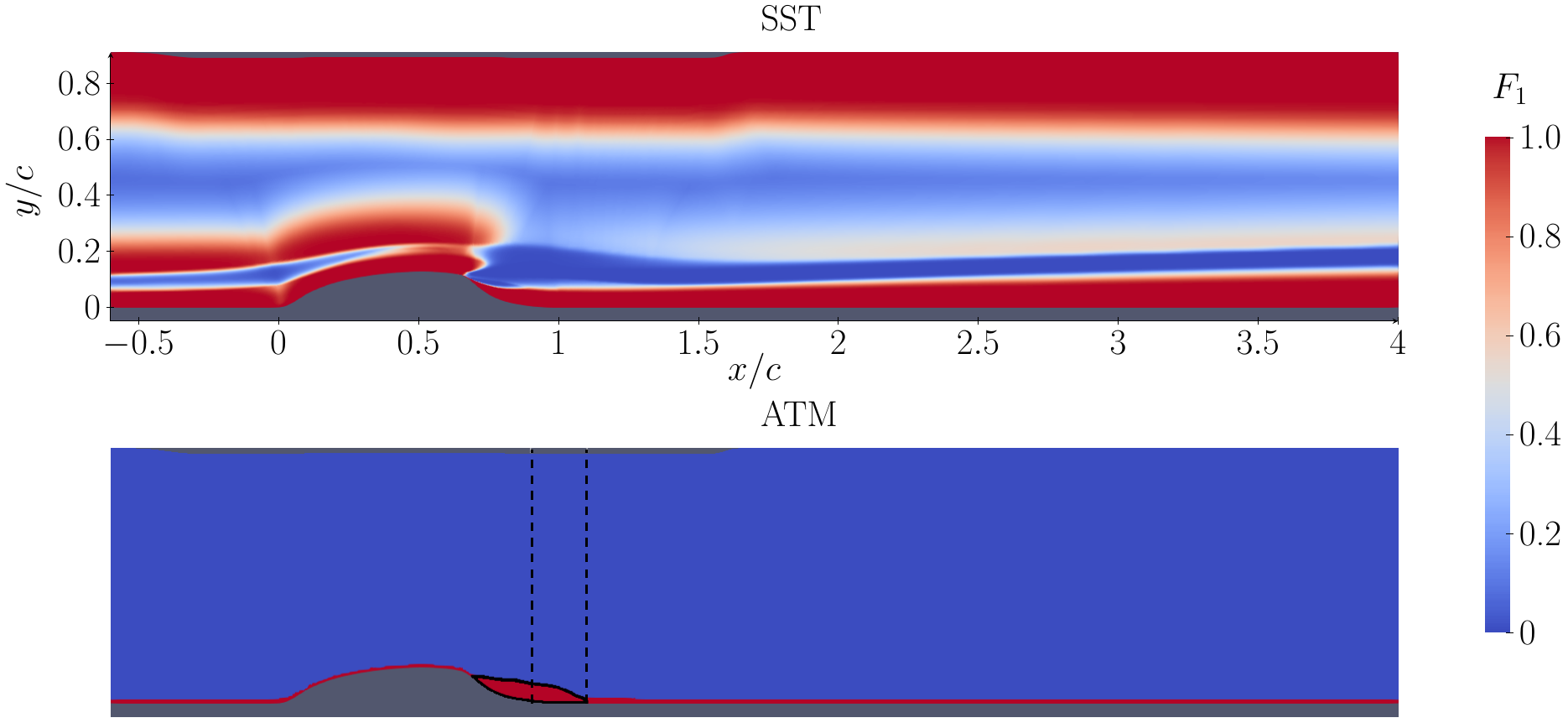}
\caption{Comparison between the empirical blending function $F_1$ of the SST model (above) and the one implicitly computed with ATM (below). $F_1=1$ corresponds to pure $k-\omega$, and $F_1=0$ corresponds to pure $k-\epsilon$. The black contour on the lower image draws the boundary of the region where $F_r=1$ and curvature correction is on.
The two vertical dashed lines mark the streamwise locations $x/c=0.9$ and $x/c=1.1$ used later in figure \ref{fig:wmhinner}.}
\label{fig:f1wmh}
\end{figure}

\begin{figure}
\centering
\includegraphics[width=\textwidth]{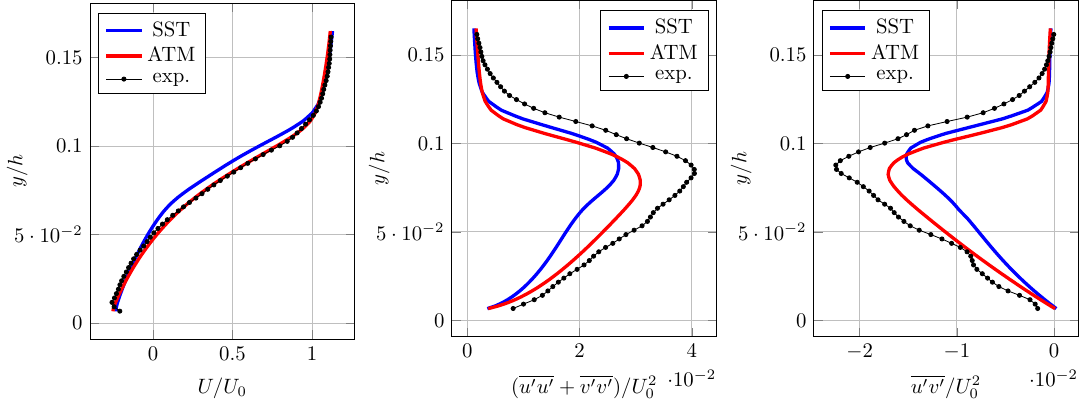}
\includegraphics[width=\textwidth]{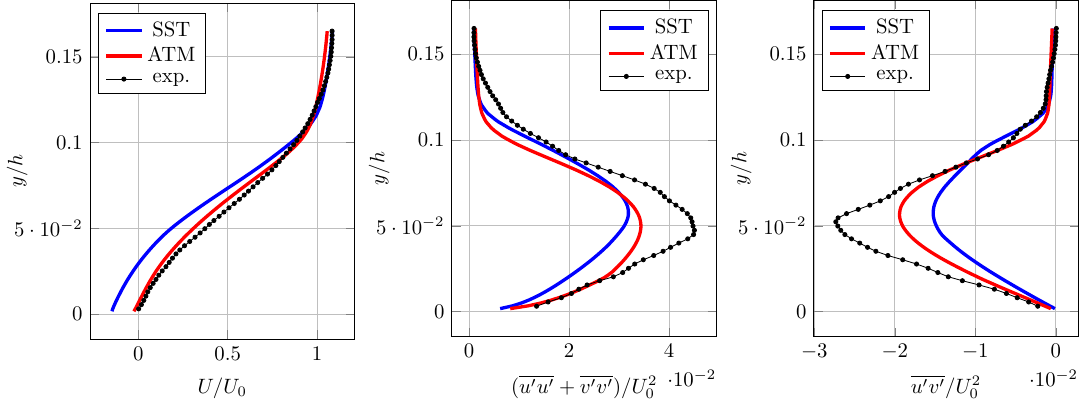}
\caption{Wall-normal profiles of the mean horizontal velocity (left), sum of the streamwise and wall-normal Reynolds stresses (center) and turbulent shear stress (right) before the end of the hump at $x/c=0.9$ (top) and after the end of the hump at $x/c=1.1$ (bottom): comparison between ATM and SST against experimental data.}
\label{fig:wmhinner}
\end{figure}

Figure \ref{fig:f1wmh} compares the different $F_1$ functions used by SST and ATM. (Note the contoured top boundary, an adjustment introduced in the original numerical setup to compensate for blockage effects in the wind tunnel experiment). 
Qualitatively the two $F_1$ fields resemble those of the previous example.
SST predominantly mandates the use of $k-\omega$, except in the bulk of the channel, where the model sits midway between the two constituent models, and in a region that starts from the hump tip and extends downstream while slowing moving outwards. The physical reason for the presence of these regions is not entirely clear. ATM, instead, prescribes a rational distribution of $F_1$. It uses by design $k-\omega$ near the wall and in the recirculating bubble (where $k-\omega$ switches to the version with curvature correction) and $k-\epsilon$ in the shear layer, just outside the recirculation. As mentioned before, $k-\epsilon$ is supposed to better capture the intensity of turbulent shear stresses in shear layers (as visible also from figure \ref{fig:wmhinner}), with a better representation of recirculation. 

\begin{figure}
\centering
\includegraphics[width=\textwidth]{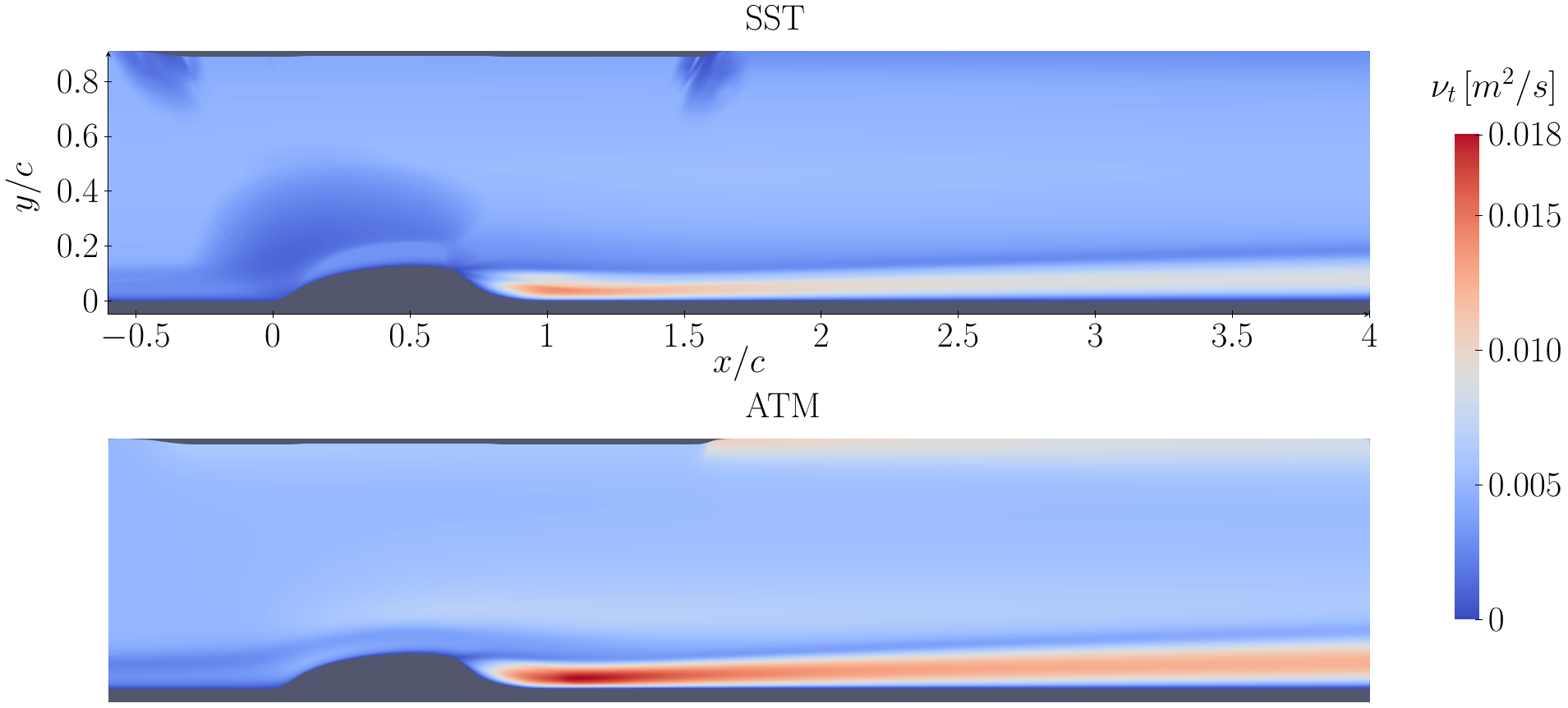}
\caption{Comparison between the field of the eddy viscosity $\nu_t$ of the SST model (above) and the one computed within the ATM procedure (below).}
\label{fig:nutwmh}
\end{figure}

Figure \ref{fig:nutwmh} compares instead the fields of turbulent viscosity computed by SST and ATM. Both fields are reasonable and qualitatively similar; ATM tends to prescribe higher values of $\nu_t$ in the wake of the hump. 
Once again, the most important takeaway is that the ATM-computed $F_1$ distribution with sharp transitions does not translate into any irregularity in the field of $\nu_t$, which is even smoother than that computed by SST.

A better understanding of the performance of the various models can be obtained by examining the wall-normal distribution of quantities measured in the experiments, namely the mean velocity profile $U$, the sum of the streamwise and wall-normal diagonal Reynolds stresses ($\overline{u'u'}+\overline{v'v'}$), and the turbulent shear stress $\overline{u'v'}$, a critical quantity that RANS models often struggle to predict correctly. 
Figure \ref{fig:wmhinner} illustrates their distribution along the vertical direction at $x/c=0.9$ (just before the end of the WMH) and $x/c=1.1$ (after the end of the WMH, at the verge of reattachment), where experimental data are available.
In both streamwise locations ATM results for $U$ align more closely to experimental data than SST, indicating a superior ability of ATM to capture the correct dimension of the recirculating bubble. Additionally, the turbulent kinetic energy and turbulent shear stresses predicted by ATM are generally larger, and in better agreement with experimental observations.

\subsection{Robustness and computational cost}
\label{sec:verification}

We conclude the presentation of the ATM approach by discussing its robustness, as well as its implications in terms of computational cost, which turn out to be negligible. The discussion will only consider the first example presented above in \S\ref{sec:step}.

First of all, the solution obtained at convergence by the RANS solver is verified to be independent from the initial specification of the RANS models. 
Indeed, even a fully implemented ATM will see the start of the iterative solution to rely on an initial condition for the spatial distribution of the models.
For this verification, two simulations are run, starting from two uniform spatial distributions of $F_1$: $F_1=0$ (equivalent to a $k-\epsilon$ initialization) and $F_1=1$ (equivalent to a $k-\omega$ initialization). 
Although the initial phases of the convergence process obviously differ, both simulations quickly converge to the same solution, in terms of both the final spatial distribution of $F_1$ and of the computed flow field. 

\begin{figure}
\centering
\includegraphics[height=0.17\textheight]{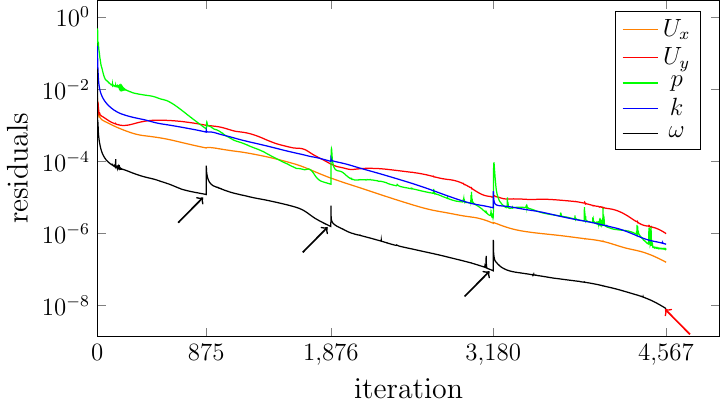}
\includegraphics[height=0.17\textheight]{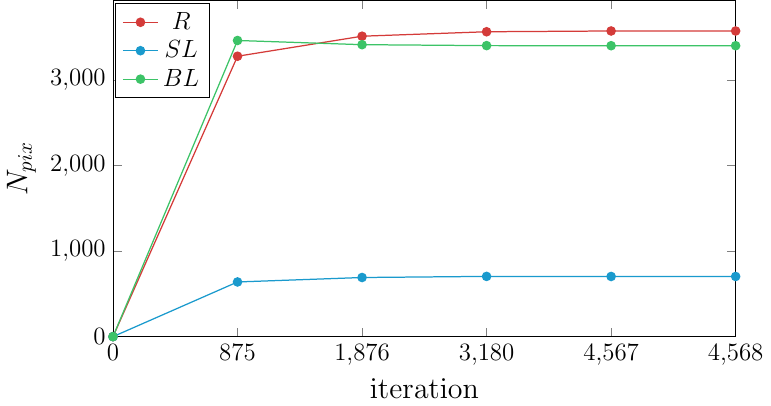}
\caption{Convergence history for the residuals (left) and the segmentation mask (right). Left: residuals of the flow and turbulence variables versus iteration number. The black arrows highlight the segmentation update by the NN; the red arrow at the end indicates the last update, without spikes. Right: number of pixels $N_{pix}$ assigned to the three labels {\em R}, {\em BL} and {\em SL}.}
\label{fig:residuals}
\end{figure}

Next we verify that the NN segmentation and the RANS solution achieve simultaneous convergence. The convergence history of the residuals of the equations for the velocity components and the turbulence variables is shown in figure \ref{fig:residuals} (left).
The simulation is set to update the spatial distribution of $F_1$ and $F_r$, and thus the layout of turbulence models, as soon as the largest residual drops below $10^{-3}$; afterwards, an update is run whenever a drop by one additional order of magnitude is observed; the iterative solution advances until the largest residual reaches $10^{-6}$. 
The arrows in figure \ref{fig:residuals} (left) highlight the update runs which cause a new segmentation. The residuals of pressure and the turbulence variables, especially $\omega$, present an evident spike occurring at the iteration where the NN recomputes the labels and subsequently updates the spatial distribution of $F_1$ and $F_r$; velocities appear to be less affected. Owing to the logarithmic vertical scale employed by the plot, the amplitude of the spikes remains visible as the simulation advances until final convergence. At convergence, the NN is run one last time (as shown by the red arrow at the end of convergence history). The lack of any spike there confirms full convergence of the solution.
Figure \ref{fig:residuals} (right) examines the convergence history in terms of number of pixels $N_{pix}$ assigned to labels {\em R}, {\em BL} and {\em SL}. After the first update, the number of pixels shows little changes, and remains essentially constant after the second update.

\begin{table}
\centering
\begin{tabular}{cccccc}     
 & ATM & \sout{\em SL} & \sout{\em R} & \sout{\em BL} & \sout{\em SL, R, BL}  \\ 
\hline
$\%\Delta X^{(1)}_r$ & $+3.3$ & $+15.6$ & $-1.5$  & $-6.3$  & $-1.5$  \\ 
$\%\Delta X^{(2)}_r$ & $-4.6$ & $+8.4$  & $-34.4$ & $-11.6$ & $-32.3$ \\
$\%\Delta Y^{(2)}_r$ & $-5.1$ & $-8.7$  & $-28.8$ & $-5.6$  & $-28.9$ \\
\end{tabular}
\caption{Length $X_r$ and width $Y_r$ of the primary and secondary recirculating bubbles for the backward-facing step flow: percentage difference between RANS and DNS. Comparison between standard ATM and ATM with incorrect label/model pairing.}
\label{tab:bfsopposto}
\end{table}

Since the ATM in its present, elementary form ends up being just a combination of three constituent models, each of whom already provides a decent outcome out of the box, the encouraging results described above could be just a lucky coincidence, instead of descending from the correct pairing between a turbulence model and the corresponding label. 
To rule out this possibility, two further experiments are carried out, in which  the validity of our choice for the $F_1$ and $F_r$ functions are tested separately. 
For the first experiment, three additional simulations (indicated with \sout{\em SL}, \sout{\em R} and \sout{\em BL} in table \ref{tab:bfsopposto}) are carried out, in which one label at a time is assigned the inverted value $1-F_1$ instead of $F_1$.
One last simulation, indicated with \sout{\em SL,R,BL}, has the pairing between labels and $F_1$ fully inverted. 
Results of these numerical experiments are shown in table \ref{tab:bfsopposto}, and confirm that the best result overall in terms of dimensions of the predicted recirculating regions  is achieved by the correct ATM, thus attesting the importance of improving the representation of the flow physics via correct zonal modelling.

\begin{figure}
\centering
\includegraphics[width=0.8\textwidth]{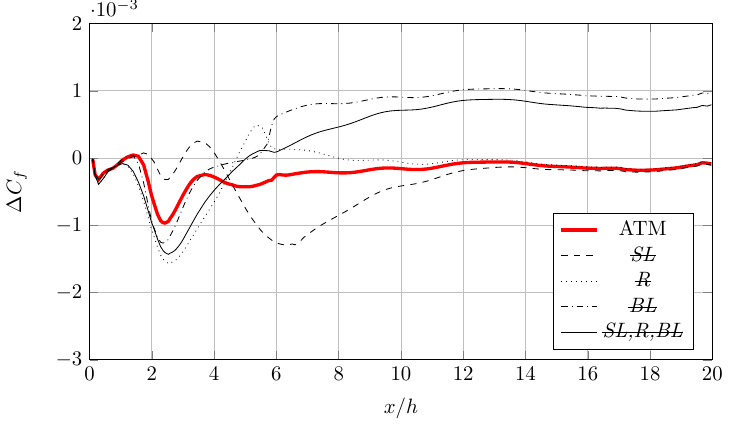}
\caption{Error $\Delta C_f$ versus DNS for the prediction of the friction coefficient along the lower wall: comparison between ATM and its variants with incorrect model-label pairing.}
\label{fig:cfopposto}
\end{figure}

The same message is extracted from figure \ref{fig:cfopposto}, that plots the error $\Delta C_f$ (computed against the DNS data) in the evolution of the skin-friction coefficient along the wall. 
In the first part of the recirculation bubble, \sout{\em SL} works well but misses the length of the recirculation; \sout{\em R} and \sout{\em SL,R,BL} represent quite well the length of the recirculation but yield poor performances in the first region of the recirculating bubble and in the recovery region, where \sout{\em BL} is also off.

For the second experiment, one additional simulation is carried out where the curvature correction term is switched off by setting $F_r=0$ for label {\em R}. 
The model without the curvature correction still performs quite well and overpredicts the reattachment length by 2\% only, but the representation of the secondary bubble gets worse, with $\Delta X^{(2)}_r$ deteriorating to $-14\%$, compared to $\Delta X^{(2)}_r = -5\%$ with the original model.

\begin{table}
\centering
\begin{tabular}{ccccc}
           & $k-\epsilon$ & $k-\omega$ & SST  & ATM   \\ 
\hline 
iterations & 5067         & 3688       & 3807 & 4567   \\ 
\end{tabular}
\caption{Iterations to convergence for the BFS example.}
\label{tab:iterations}
\end{table}

A final, important consideration concerns the computational cost of the ATM approach. 
Although computational cost is nowadays not the primary concern for RANS simulation, if ATM incurred in a significant cost aggravation then the comparisons carried out so far would not be entirely fair, and a comparison at the same cost would be in order.
Luckily, the computational overhead of ATM is negligible. 
In fact, segmentation does not need to take place continuously during the convergence process. The procedure employed here, albeit empirically devised and thus probably less than optimal, has shown that a few calls to the segmentation network during the run are sufficient, and the cost of such calls is negligible. 
Even the convergence rate is not negatively affected by ATM, as shown in table \ref{tab:iterations}, reporting the number of iterations needed to reach the end of the iterative solution, conventionally defined at the iteration when the largest residual shrinks below $10^{-6}$. 
As expected, against the improved predictive performance, the number of iterations to convergence remains intermediate between that of $k-\epsilon$ and $k-\omega$.

\section{Concluding discussion}
\label{sec:conclusions}

We have introduced an approach to RANS turbulence modelling, dubbed ATM, which does not require user input. A neural network for semantic segmentation is used at runtime to extract different structural flow regions in the computed flow field; each region is assigned to a turbulence model, previously optimized for the physics of that specific flow zone, without the requirement of universality. Training the network does not require high-fidelity data, which constitutes one of the major advantages of ATM. 
A minimalistic version of the ATM has been implemented with relative ease into an open-source, industrial flow solver, and successfully demonstrated on two representative examples for which high-fidelity information is available.

The ATM procedure hinges upon a runtime semantic segmentation of the flow field, which is interpreted as a multi-channel image. Segmentation is carried out with a modern neural network for segmentation, the U-net. 
The computational effort required to produce the training and validation dataset is next to negligible. 
Training data have been created with cheap two-dimensional laminar simulations, by parametrically varying a handful of simple flow configurations. 
In training the U-net, a customized loss function has been used, where categorical cross entropy and intersection over union are combined to create a loss function that performs well during every stage of the training.
The only relatively time-consuming part in setting up ATM is the manual annotation of the dataset to obtain the ground-truth labels. Although not attempted here, alternatives to manual annotation exist; however, it should be kept in mind that annotation is a one-time task.
Once trained, the neural network has provided fully satisfactory segmentation results on several test cases, including laminar and turbulent flows as well, thanks to the qualitative resemblance between a laminar flow field and a turbulent mean flow field.

The present reinterpretation of the zonal modelling approach differs from the one by \cite{matai-durbin-2019-2} in the ability to dispense with high-fidelity training data, which is a crucial problem in ML-based turbulence modelling at large.
While in that work the ML model directly computes a zonal coefficient, our approach relies on the ML model only to identify the structural flow zones, separating this step from the choice of the associated turbulence model.
Similarly, the ATM presented here is, in some respects, similar to the approach introduced by \cite{dezordo-etal-2024}, where solutions derived from different RANS models are linearly combined in the attempt to locally give preference to the model that works best. The random-forest algorithm that drives the combination, however, still requires high-fidelity data for training. 

The predictive capabilities of the ATM have been put to the test on two flow cases that are often used in the assessment of RANS models: the flow over a backward-facing step, for which the DNS by \cite{le-moin-kim-1997} in an open-channel configuration is available, and the flow over a wall-mounted hump, where the reference is the experiment by \cite{greenblatt-etal-2006}. 
Our implementation of ATM, albeit preliminary, already outperforms both the $k-\epsilon$ and $k-\omega$ models, and is equivalent to or better than the $k-\omega$ SST model, which is often considered the gold standard.

Galilean invariance is an essential property of RANS models. Since a combination of invariant models is invariant as well, enforcing this property in ATM only requires the design of a Galilean-invariant segmentation. 
Indeed, in the present work the training set consists of images where the flow is described in a reference system where the solid body is at rest. 
However, a different inertial reference frame can be accounted for during training: once the speed of the body is known, the network can subtract it such that the body becomes at rest.

In view of the important practical issues mentioned by \cite{spalart-2023} regarding the publishability of a ML model, the ATM approach seems to tick most if not all the important boxes. 
The integration of the NN within an open-source solver is relatively straightforward, provided the constituent models are available, and becomes very simple if the same model with variable coefficients is used, or a model setting with free functions is employed.
The NN used here has enough degrees of freedom for a full two-dimensional implementation with, say, ten flow zones; publishing the trained version and using it in a RANS solver requires a disk/memory space of about 60 MB, which includes both the NN structure and the set of coefficients optimized after training. 
In the spirit of openness, sharing the training set on top of the NN would be useful for those wishing to extend or customize the ATM; the present set of images used for training only weighs few megabytes, hinting at a very reasonable size for the training set of the full-scale version of the model.

The implementation of the ATM concept is not unique. 
Besides, the presented work is just a proof of concept, and is certainly not enough to asses the full potential of ATM: the obtained results should be considered merely as a starting point.
Among the various stages of the procedure which need to be thoughtfully designed and implemented, three aspects stand out for their importance: i) the choice of the models to include (starting from the preliminary decision whether several models are used in the procedure, or a single model is used with multiple sets of values for the model constants corresponding to the various zones); ii) the number and type of elementary flows considered in the dictionary for the semantic segmentation (the present value of three is certainly insufficient, but going up to the 20 zones listed by \cite{kline-1980} may be unnecessary); iii) the type, size and quality of the dataset used for training (including tools and techniques to avoid manual annotations).
All these limitations notwithstanding, the workflow has proven to be feasible and reliable in terms of segmentation accuracy, and the obtained results are already more than satisfactory.  
We have also discussed how ATM is robust with respect to several aspects of the procedure, and that it does not incur in additional computational costs.

At least in theory, ATM holds the potential to introduce a significant change in the way industrial CFD simulations deal with turbulence modelling, because it promises to dispense the average user with explicitly picking a specific turbulence model. Instead, the informed design choices made by the turbulence modelling community would back up the automatic selection of the specific model to adopt in a certain zone of the flow.
However, bringing out the true potential of ATM is not going to be immediate. It requires increasing the number of flow labels, exploring the best strategy to associate them to turbulence models, and specializing a set of models towards the chosen labels. 
Such a task is well outside the scope of the present work, and will require a community effort in which the turbulence modelling experts will be called to provide significant input.

\section*{Acknowledgments}
Computational resources have been provided by CINECA via the ISCRA C project ETMNN. The authors acknowledge early discussions with prof. G.Boracchi and Dr A.Chiarini, who also helped in setting up the FreeFEM simulations. We are grateful for useful discussions on some aspects of this work with Profs. A.Abb\`a and G.Persico.

\bibliographystyle{jfm}

\end{document}